\documentstyle[12pt]{article}

\font\mybb=msbm10 at 10pt
\def\bb#1{\hbox{\mybb#1}}
\def\Z {\bb{Z}}
\def\R {\bb{R}}
\def\unit{\hbox to 3.3pt{\hskip1.3pt \vrule height 7pt width .4pt \hskip.7pt
\vrule height 7.85pt width .4pt \kern-2.4pt
\hrulefill \kern-3pt
\raise 4pt\hbox{\char'40}}}
\def\II{{\unit}}

\makeatletter
\def\setboxz@h{\setbox\z@\hbox}
\def\binrel@#1{\begingroup
  \setboxz@h{\thinmuskip0mu
    \medmuskip\m@ne mu\thickmuskip\@ne mu
    \setbox\tw@\hbox{$#1\m@th$}\kern-\wd\tw@
    ${}#1{}\m@th$}%
  \edef\@tempa{\endgroup\let\noexpand\binrel@@
    \ifdim\wd\z@<\z@ \mathbin
    \else\ifdim\wd\z@>\z@ \mathrel
    \else \relax\fi\fi}%
  \@tempa
}
\let\binrel@@\relax
\def\overset#1#2{\binrel@{#2}%
  \binrel@@{\mathop{\kern\z@#2}\limits^{#1}}}
\makeatother

\begin{document}
\begin{titlepage}

\begin{flushright}
UG--3/95\\
QMW--PH--95--2\\
{\bf hep-th/9504081}\\
April $16$th, $1995$\\
Revised October $12$th $1995$
\end{flushright}

\begin{center}

\baselineskip20pt
{\LARGE {\bf DUALITY IN THE TYPE--II SUPERSTRING EFFECTIVE ACTION}}

\vspace{.4cm}

\baselineskip15pt

{\bf Eric Bergshoeff}\\
{\it Institute for Theoretical Physics, University of Groningen}\\
{\it Nijenborgh 4, 9747 AG Groningen, The Netherlands}\\
\vspace{.4cm}
{\bf Chris  Hull}
{\bf and  Tom\'as Ort\'{\i}n}
\footnote{Address after October $1^{st}$: Theory Division, C.E.R.N.,
CH-1211, Gen\`eve 23, Switzerland.}\\
{\it Department of Physics, Queen Mary and Westfield College}\\
{\it Mile End Road, London E1 4NS, U.K.}

\end{center}

\vspace{.4cm}


\begin{abstract}

We derive the $T$--duality transformations that transform a general
$d=10$ solution of the type--IIA string with one isometry to a solution
of the type--IIB string with one isometry and vice versa.  In contrast
to other superstring theories, the $T$--duality transformations are not
related to a non-compact symmetry of a $d=9$ supergravity theory.  We
also discuss $S$--duality in $d=9$ and $d=10$ and the relationship with
ele\-ven-di\-men\-sion\-al supergravity theory.  We apply these
dualities to generate new solutions of the type--IIA and type--IIB
superstrings and of ele\-ven-di\-men\-sion\-al supergravity.

\end{abstract}

\end{titlepage}

\newpage

\pagestyle{plain}
\baselineskip 16pt


\section*{Introduction}

Duality symmetries \cite{kn:Gi1, kn:Sen1, kn:Hu1, kn:Du1, kn:Ma1} play
an important role in string theories and it has recently been found that
duality symmetries of type--II strings have a number of interesting and
unusual features \cite{kn:Hu1}.  The aim of this paper is to explore
duality symmetries and some of their applications in the context of the
type--II string in nine and ten dimensions, and the relation of these to
ele\-ven-di\-men\-sion\-al supergravity.  In particular, we aim to
understand the $T$--duality symmetry of the type--II string in
backgrounds with one isometry.  This symmetry is of a rather unusual
type in that it maps type--IIA backgrounds into type--IIB ones, and vice
versa \cite{kn:Da1,kn:Di1}.  Moreover, whereas in the heterotic string
$T$--duality for backgrounds with one isometry can be understood as a
symmetry of ni\-ne-di\-men\-sion\-al $N=1$ supergravity, no such
understanding is possible here: the type--II $T$--duality does not
correspond to any symmetry of the ni\-ne-di\-men\-sion\-al $N=2$
supergravity theory.  A discussion of our results has been given
recently by one of us \cite{kn:Be1}, and there is some overlap with the
results of Witten \cite{kn:Wi1} announced at the same conference.

The bosonic string compactified from $D+d$ dimensions to $D$ dimensions
on a $d$--torus $T^d$ has an $O(d,d)$ duality symmetry which is broken
to the discrete subgroup $O(d,d;\Z)$ by non-perturbative sigma-model
effects.  (Either $D+d=26$, or there is an additional hidden sector
describing internal degrees of freedom through a CFT with $c=26-D-d$,
which is suppressed in the following.)  This discrete target-space
duality or $T$--duality group includes the well-known $R \rightarrow
\alpha^\prime /R$--duality for each circle in $T^d$, where $R$ is the
radius, together with shifts of the antisymmetric tensor gauge field and
$O(d;\Z)$ rotations of the circles into one another; the latter are
particular $D+d$ dimensional diffeomorphisms.  The $O(d,d;\Z)$ is a
discrete gauge group, and configurations related by such a duality
transformation are physically equivalent.  The $O(d,d)$ group is not a
string symmetry, but transforms a consistent string background to a new
one.

This can be generalized to consider the string on a curved $D+d$
dimensional space with $d$ commuting isometries.  For consistency, the
background must define a conformally invariant sigma-model which implies
that the background fields must satisfy certain field equations, which
can be derived from a low-energy effective action.  There is again an
$O(d,d)$ symmetry transforming solutions of the low-energy equations of
motion into new ones; this was first shown for $d=1$ by Buscher
\cite{kn:Bu1} and generalized to higher $d$ in Ref.~\cite{kn:Ro}.  If
the orbits of the isometries are compact, then there is again a discrete
subgroup $O(d,d;\Z)$ which is a discrete gauge symmetry of the string
theory, but there will be Buscher duality even for non-compact orbits.
Although the name $T$--duality is usually reserved for the discrete
group $O(d,d;\Z)$, we shall refer here also to the $O(d,d)$ Buscher
group as a $T$--duality.

For the heterotic superstring without background gauge fields and the
type--II superstring without background fields from the Ramond-Ramond
(RR) sector, the situation is similar.  For such backgrounds with $d$
commuting isometries, the arguments of Refs.~\cite{kn:Bu1, kn:Ro} give
an $O(d,d)$ $T$--duality symmetry of the equations of motion and an
$O(d,d;\Z)$ discrete gauge symmetry if the orbits are compact.  For the
heterotic string, including sixteen background Abelian gauge fields
enlarges the $T$--duality group to $O(d,d+16)$.  The main aim here will
be to study duality in the type--II string in the presence of background
RR fields.  In particular, we shall be interested in
ten-di\-men\-sion\-al type--II backgrounds with one isometry.  For a
string moving in the special background $M^{9} \times T^{1}$
(ni\-ne-di\-men\-sion\-al Minkowski space times a one--torus) this has
already been done in Refs.~\cite{kn:Da1,kn:Di1}.  In these references it
was argued that there is a $\Z_2$ $T$--duality symmetry that relates the
type--IIA string on a circle of radius $R$ and type--IIB string moving
on a circle of radius $ \alpha^\prime /R$.  We will find a
generalization of Buscher's transformation that transforms a solution of
the type--IIA string on a background with one isometry to a solution of
the type--IIB string on a background with one isometry.  The
transformation is essentially that of Buscher when restricted to fields
in the Neveu-Schwarz/Neveu-Schwarz (NS-NS) sector, but interchanges the
RR background fields of the type--IIA string with those of the type--IIB
string.  We shall use such transformations to generate new solutions of
the equations of motion.  Since this type--II $T$--duality maps all the
fields of the type--IIA string into the type--IIB and vive versa, it
also maps the symmetries and can be used, for instance, to find in the
type--IIA theory with one isometry the form of the $SL(2,\R)$--duality
transformations which are well-known in the type--IIB theory.

Duality symmetries of the full string theory necessarily give rise to
symmetries of the low-energy effective supergravity theory.  In this
paper, we shall study duality symmetries of such effective supergravity
theories to lowest order in $ \alpha '$, which constitutes a first step
toward studying string dualities.  In addition to the $T$--dualities
which are perturbative string symmetries and so can be studied using
world-sheet sigma-models, there are also symmetries of the supergravity
actions which correspond to conjectured non-perturbative $S$-- and $U$--
duality string symmetries.  In particular, the $S$--duality group
includes transformations which act on the dilaton $\hat \phi$.  The
type--IIB supergravity in ten dimensions has an explicit $SL(2,\R)$
symmetry of the equations of motion which is broken to $SL(2,\Z)$ by
quantum corrections.  This is the conjectured $SL(2,\Z)$ $S$--duality
discrete gauge symmetry of the type--IIB string \cite{kn:Hu1}, while
$SL(2,\R)$, which we shall also refer to as an $S$--duality, is a
solution-generating symmetry, transforming any given solution into a new
one.  The type--IIA has an $SO(1,1)$ $S$--duality symmetry which acts
rather trivially through a shift of the dilaton and scaling of the other
fields.  There is only one kind of $N=2$ supergravity in nine
dimensions, so that compactifying either type--IIA or type--IIB
supergravities to nine (or less) dimensions gives the same compactified
theory, which inherits the symmetries of both of its two parent
theories.  The ni\-ne-di\-men\-sion\-al theory has an $SL(2,\R) $
symmetry, which is broken to the $SL(2,\Z)$ $S$--duality by quantum
effects \cite{kn:Hu1}.  The $SL(2,\R)$ can be thought of as arising from
the $SL(2,\R)$ symmetry of the type--IIB theory, but its origins from
the type--IIA theory are not so clear.  In \cite{kn:To1} a relation
between the type--IIA string and the 11-dimensional membrane
compactified to d=10 on a circle was suggested in which the dilaton
emerges as a modulus field for the compact dimension.  We shall find
further evidence for the role of 11-dimensions in the type--IIA string.
In particular, we show that some of the $SL(2,\R)$ duality of the
type--IIA theory compactified to d=9 has a natural interpretation in
eleven dimensions: an $O(2)$ subgroup of $SL(2,\R)$ can be interpreted
as ele\-ven-di\-men\-sion\-al Lorentz transformations.  The relevance of
ele\-ven-di\-men\-sion\-al supergravity to the type--IIA string has been
discussed independently by Witten \cite{kn:Wi1}.

The ni\-ne-di\-men\-sion\-al theory is also expected to have an $O(1,1)$
or $SO(1,1)$ symmetry which is related to $T$--duality.  Indeed, when
truncated to the NS-NS sector, the ni\-ne-di\-men\-sion\-al theory
indeed has an $O(1,1)=SO(1,1)\times \Z_2$ symmetry which has a $\Z _2$
subgroup that corresponds to the expected $R \to \alpha ' /R$
$T$--duality.  However, only an $SO(1,1)$ subgroup extends to a symmetry
of the full $N=2$ theory in nine dimensions, while the $\Z _2$ ``$R \to
\alpha ' /R$" duality does not correspond to any such symmetry of the
$d=9$ supergravity.  One of our aims is to elucidate the extension of
this $\Z_2$ symmetry to the type--II theory and show how it can be
understood in terms of supergravity theories.

We shall investigate the extent to which these type--II dualities can be
interpreted as symmetries of $d=10$ theories on backgrounds with one
isometry and of $d=11$ theories on backgrounds with two isometries.  As
has already been mentioned, the $T$--duality gives a solution-generating
transformation which takes type--IIA to type--IIB, and vice-versa.  As
the type--IIB theory has no known ele\-ven-di\-men\-sion\-al origin, we
will only be able to lift the ni\-ne-di\-men\-sion\-al dualities to
solution--generating transformations of $d=11$ supergravity on
backgrounds with two isometries for a special restricted class of
backgrounds satisfying certain geometric and algebraic conditions.
Eleven-dimensional supergravity is the low-energy limit of the
ele\-ven-di\-men\-sion\-al supermembrane \cite{kn:Be2}; a search for
supermembrane duality symmetries was undertaken in the context of a
(three-dimensional) sigma-model description of supermembranes in
Refs.~\cite{kn:Du1,kn:Se1}.  We are able, in addition, to explicitly
construct another set of solution--generating transformations that acts
only inside each of the type--II strings on backgrounds with one
isometry, behaves as a strong--weak coupling duality and is therefore
part of $SL(2,\R)$.

As an illustration of how our results can be applied to generate new
solutions to the string equations of motion, we will consider in this
paper the Supersymmetric String Wave (SSW) solution of
Ref.~\cite{kn:Be3} which is a solution of the heterotic string but also
solves the type--II string equations of motion.  Under a type--I
$T$--duality transformation the SSW solution generates the Generalized
Fundamental String Solution of Ref.~\cite{kn:Be4} which is a
generalization of the fundamental string solution \cite{kn:Dab1}.  We
will show how the application of both type--II $S$-- and $T$--dualities
as well as combinations thereof generate new solutions of the type--IIA
and type--IIB equations of motion.  We lift the SSW solution to a
solution of the eleven--dimensional theory.  This gives a generalization
of the ele\-ven-di\-men\-sion\-al pp-wave solution constructed in
\cite{kn:Hu2}.  Applying a $d=11$ type--I $T$--duality transformation
generates an ele\-ven-di\-men\-sion\-al Generalized Fundamental Membrane
(GFM) solution which is a generalization of the fundamental membrane
solution of \cite{kn:Du2}.

In exploring these symmetries, we work out some of the details of
supergravity theories in $d=9,10$ that have not appeared in the
literature before.  We give the bosonic part of the $N=2,d=9$ action,
which has not been written down explicitly before.  We also write the
type--IIA supergravity action for the stringy metric and find that,
whereas the NS-NS fields appear with a coupling to the dilaton
$\hat\phi$ through an overall factor of $e^{-2\hat{\phi}}$, as expected,
the RR fields appear without any dilaton coupling.  This follows from
the fact that, on compactifying to four dimensions for example, the RR
fields are invariant under $S$--duality \cite{kn:Hu1} and this implies
that they cannot couple to the dilaton in a way that respects
$S$--duality; this was noticed independently by Witten
\cite{kn:Wi1}\footnote{It was already known that the
four-di\-men\-sio\-nal dilaton-axion field cannot couple to RR vectors
\cite{kn:Ce1} }.

The organization of this work is as follows.  In
Section~\ref{sec-11to10} we first derive the action of type--IIA
supergravity in the ``string-frame'' metric.  This action describes the
zero--slope limit ($\alpha^{\prime} \rightarrow 0$) of the type--IIA
superstring.  We use here dimensional reduction from eleven dimensions.
In order to derive the type--II $T$--duality rules we first reduce in
the next section the type--IIA supergravity theory to nine dimensions
and thus obtain the action of $N=2, d=9$ supergravity.  Next, in
Section~\ref{sec-IIB} we present the equations of motion of type--IIB
supergravity in the ``string-frame'' metric and discuss its reduction to
nine dimensions.  Using all this information we derive in
Section~\ref{sec-dualityin10} the explicit form of the above--mentioned
type--II $S$-- and $T$--duality rules.  In Section~\ref{sec-elevendual}
we use our results to derive a type--I $T$--duality symmetry of
$N=1,d=11$ supergravity, the only duality symmetry that can be found in
this framework.  Finally, as an illustration of how our results can be
applied to generate new solutions, we will apply in
Section~\ref{sec-examples} the duality transformations constructed in
this work to the SSW solution and generate new, ``dual'', solutions of
the type--IIA and type--IIB superstrings and of
ele\-ven-di\-men\-sion\-al supergravity.  Our conventions are explained
in Appendix~\ref{sec-conventions} and Appendix~\ref{sec-11vs9} contains
some useful formulae giving the explicit relations between certain
eleven-- and ni\-ne-di\-men\-sion\-al fields.


\section{The Type-IIA Superstring}
\label{sec-11to10}

The zero--slope limit of the type--IIA superstring corresponds to $N=2,
d=10$ non--chiral supergravity.  In this section we describe how to
obtain the (bosonic sector of) type--IIA supergravity in the
``string-frame'' metric by dimensional reduction of $N=1,d=11$
supergravity.  This can be done by a straightforward application of
standard techniques (see for instance Ref.~\cite{kn:Sc1}).  We describe
the dimensional reduction in some detail since in order to derive the
type--II duality rules (see Section~\ref{sec-dualityin10}) we need to
know the exact relation between the supergravity theories in different
dimensions.  In this paper we will describe supergravity theories in
$d=9,10$ and $11$ dimensions.  It is helpful to use a notation that
clearly distinguishes between the different dimensions; throughout this
paper we will use double hats for ele\-ven-di\-men\-sion\-al objects,
single hats for ten-di\-men\-sion\-al objects and no hats for
ni\-ne-di\-men\-sion\-al objects.

We now proceed to describe the dimensional reduction of $N=1,d=11$
supergravity \cite{kn:CJS}.  The bosonic fields of this theory are the
elfbein and a three-form potential

\begin{equation}
\left\{\hat{\hat{e}}_{\hat{\hat{\mu}}}{}^{\hat{\hat{a}}},
\hat{\hat{C}}_{\hat{\hat{\mu}}\hat{\hat{\nu}}\hat{\hat{\rho}}}
\right\}\, .
\end{equation}

The field strength of the three-form is

\begin{equation}
\hat{\hat{G}} =\partial \hat{\hat{C}}\, ,
\end{equation}

\noindent and the action for these bosonic fields is\footnote{For
simplicity, we have set the fermions to zero.  It is straightforward to
include fermion fields in the following analysis.}

\begin{equation}
\hat{\hat{S}}= {\textstyle\frac{1}{2}}\int d^{11}x\
\sqrt{\hat{\hat{g}}}\ \left[-\hat{\hat{R}} +a_{1}\hat{\hat{G}}^{2}
+a_{2}\frac{1}{\sqrt{\hat{\hat{g}}}}
\hat{\hat{\epsilon}} \hat{\hat{G}} \hat{\hat{G}} \hat{\hat{C}}
\right]\, .
\label{eq:11daction}
\end{equation}

\noindent We use the index-free notation explained in
Appendix~\ref{sec-conventions}.  The coefficients $a_{1}$ and $a_{2}$
are numerical constants which are defined up to redefinitions of
$\hat{\hat{C}}$, which implies that only the following quotient can be
fixed:

\begin{equation}
a_{1}^{3}/a_{2}^{2}=9 (4!)^{3}/2\, .
\end{equation}

\noindent The action above is invariant under general coordinate
transformations and the following gauge transformations of the
$\hat{\hat{C}}$ potential:

\begin{equation}
\delta \hat{\hat{C}}=\partial\hat{\hat{\chi}}\, .
\label{eq:gauge11}
\end{equation}

We assume that all fields are independent of the coordinate $y
=x^{\underline{10}}$ which we choose to correspond to a space-like
direction ($\hat{\hat{\eta}}_{\underline{y}\underline{y}} =-1$) and we
rewrite the fields and action in a ten-di\-men\-sion\-al form.  The
dimensional reduction of the metric gives rise to the
ten-di\-men\-sion\-al metric, a vector field and a scalar (the dilaton)
while the dimensional reduction of the three-form potential gives rise
to a ten-di\-men\-sion\-al three-form and a two-form.  We thus obtain
the fields of the ten-di\-men\-sion\-al type--IIA supergravity theory
which are

\begin{equation}
\left\{
\hat{C}_{\hat{\mu}\hat{\nu}\hat{\rho}},
\hat{g}_{\hat{\mu}\hat{\nu}},
\hat{B}^{(1)}_{\hat{\mu}\hat{\nu}},
\hat{A}^{(1)}_{\hat{\mu}},\hat{\phi}
\right\}\, .
\end{equation}

The ele\-ven-di\-men\-sion\-al fields can be expressed in terms of the
ten-di\-men\-sion\-al ones as follows

\begin{equation}
\begin{array}{rclrcl}
\hat{\hat{g}}_{\hat{\mu}\hat{\nu}}
&
=
&
e^{-\frac{2}{3}\hat{\phi}}\hat{g}_{\hat{\mu}\hat{\nu}}
-e^{\frac{4}{3}\hat{\phi}}\hat{A}^{(1)}_{\hat{\mu}}
\hat{A}^{(1)}_{\hat{\nu}}\, ,\hspace{1cm}
&
\hat{\hat{C}}_{\hat{\mu}\hat{\nu}\hat{\rho}}
&
=
&
\hat{C}_{\hat{\mu}\hat{\nu}\hat{\rho}}\, ,
\\
& & & & &
\\
\hat{\hat{g}}_{\hat{\mu}\underline{y}}
&
=
&
-e^{\frac{4}{3}\hat{\phi}}\hat{A}^{(1)}_{\hat{\mu}}\, ,
&
\hat{\hat{C}}_{\hat{\mu}\hat{\nu}\underline{y}}
&
=
&
\frac{2}{3}\hat{B}^{(1)}_{\hat{\mu}\hat{\nu}}\, ,
\\
& & & & &
\\
\hat{\hat{g}}_{\underline{y}\underline{y}}
&
=
&
-e^{\frac{4}{3}\hat{\phi}}\, .
& & &
\\
\end{array}
\end{equation}

\noindent For the vielbeins we have

\begin{equation}
\left( \hat{\hat{e}}_{\hat{\hat{\mu}}}{}^{\hat{\hat{a}}} \right)=
\left(
\begin{array}{cc}
e^{-\frac{1}{3}\hat{\phi}} \hat{e}_{\hat{\mu}}{}^{\hat{a}}
&
e^{\frac{2}{3}\hat{\phi}} \hat{A}^{(1)}_{\hat{\mu}}
\\
&
\\
0
&
e^{\frac{2}{3}\hat{\phi}}
\\
\end{array}
\right)
\, ,
\hspace{1cm}
\left( \hat{\hat{e}}_{\hat{\hat{a}}}{}^{\hat{\hat{\mu}}} \right)=
\left(
\begin{array}{cc}
e^{\frac{1}{3}\hat{\phi}} \hat{e}_{\hat{a}}{}^{\hat{\mu}}
&
-e^{\frac{1}{3}\hat{\phi}} \hat{A}^{(1)}_{\hat{a}}
\\
&
\\
0
&
e^{-\frac{2}{3}\hat{\phi}}
\\
\end{array}
\right)\, .
\label{eq:basis}
\end{equation}

\noindent Conversely, the ten-di\-men\-sion\-al fields can be expressed
in terms of the ele\-ven-di\-men\-sion\-al ones via:

\begin{equation}
\begin{array}{rclrcl}
\hat{g}_{\hat{\mu}\hat{\nu}}
&
=
&
\left( -\hat{\hat{g}}_{\underline{y}\underline{y}} \right)^{\frac{1}{2}}
\left (\hat{\hat{g}}_{\hat{\mu}\hat{\nu}}
-\hat{\hat{g}}_{\hat{\mu}\underline{y}}
\hat{\hat{g}}_{\hat{\nu}\underline{y}}
/\hat{\hat{g}}_{\underline{y}\underline{y}} \right)\, ,\hspace{1cm}
&
\hat{C}_{\hat{\mu}\hat{\nu}\hat{\rho}}
&
=
&
\hat{\hat{C}}_{\hat{\mu}\hat{\nu}\hat{\rho}}\, ,
\\
& & & & &
\\
\hat{A}^{(1)}_{\hat{\mu}}
&
=
&
\hat{\hat{g}}_{\hat{\mu}\underline{y}}
/\hat{\hat{g}}_{\underline{y}\underline{y}}\, ,
&
\hat{B}^{(1)}_{\hat{\mu}\hat{\nu}}
&
=
&
\frac{3}{2}\hat{\hat{C}}_{\hat{\mu}\hat{\nu}\underline{y}}\, ,
\\
& & & & &
\\
\hat{\phi}
&
=
&
\frac{3}{4}\log{\left( -\hat{\hat{g}}_{\underline{y}\underline{y}}
\right)}\, .
& & &
\\
\end{array}
\label{eq:10ddef}
\end{equation}

The ten-di\-men\-sion\-al fields have been defined in this way because,
as we will see, (i) their gauge transformations are natural (no scalars
are involved) and of a standard form (see below) and (ii) if we truncate
the theory by setting $\hat{C} =\hat{A}^{(1)} =0$ we recover the bosonic
action of $N=1,d=10$ (type--I) supergravity written with the usual
conventions in the ``string-frame'' metric.

We now consider the reduction of the action Eq.~(\ref{eq:11daction}) in
more detail.  We first consider the Ricci scalar term.  To reduce this
term we use (a slight generalization of) Palatini's
identity\footnote{Since the identity is valid in arbitrary $d$
dimensions we do not use any hats here.}:

\begin{eqnarray}
\lefteqn{
\int d^{d}x\ \sqrt{|g|}\, e^{-2\phi}\ [-R]=}
\nonumber \\
& &
\nonumber \\
& &
\int d^{d}x \ \sqrt{|g|}\ e^{-2\phi}\, \left[
\omega_{b}{}^{ba}\omega_{c}{}^{c}{}_{a} +\omega_{a}{}^{bc}
\omega_{bc}{}^{a} +4\omega_{b}{}^{ba}(\partial_a\phi)
\right]\, .
\label{eq:Pal}
\end{eqnarray}

The non-vanishing components of the spin connection are are

\begin{equation}
\begin{array}{rclrcl}
\hat{\hat{\omega}}_{y\hat{a}y}
&
=
&
-\frac{2}{3} e^{\frac{1}{3}\hat{\phi}} \partial_{\hat{a}}\hat{\phi}\, ,
\hspace{1cm}
&
\hat{\hat{\omega}}_{y\hat{a}\hat{b}}
&
=
&
-\frac{1}{2} e^{\frac{4}{3} \hat{\phi}}
\hat{F}_{\hat{a}\hat{b}}^{(1)}\, ,
\\
& & & & &
\\
\hat{\hat{\omega}}_{\hat{a}\hat{b}y}
&
=
&
\frac{1}{2} e^{\frac{4}{3}\hat{\phi}} \hat{F}_{\hat{a}\hat{b}}^{(1)}\, ,
&
\hat{\hat{\omega}}_{\hat{a}\hat{b}\hat{c}}
&
=
&
e^{\frac{1}{3}\hat{\phi}} \left( \hat{\omega}_{\hat{a}\hat{b}\hat{c}}
+\frac{2}{3} \delta_{\hat{a}[\hat{b}} \partial_{\hat{c}]}
\hat{\phi} \right)\, ,
\\
\end{array}
\end{equation}

\noindent where

\begin{equation}
\hat{F}^{(1)} =2\partial \hat{A}^{(1)}\,
\end{equation}

\noindent is the field strength of the ten-di\-men\-sion\-al vector
field $\hat{A}^{(1)}_{\hat{\mu}}$.  Ignoring the integration over $y$
and using

\begin{equation}
\sqrt{\hat{\hat{g}}}\ =\sqrt{-\hat{g}}\ e^{-\frac{8}{3}\hat{\phi}}\, ,
\end{equation}

\noindent plus Palatini's identity Eq.~(\ref{eq:Pal}) for $d=11$ and
$\phi=0$ we find

\begin{eqnarray}
\lefteqn{{\textstyle\frac{1}{2}}\int d^{11}x\
\sqrt{\hat{\hat{g}}}\ [-\hat{\hat{R}}]= }
\nonumber \\
& &
\nonumber \\
& &
{\textstyle\frac{1}{2}}
\int d^{10}x\ \sqrt{-\hat{g}} \left\{ e^{-2\hat{\phi}}
\left[ \left( \hat{\omega}_{\hat{b}}{}^{\hat{b}\hat{a}}
+2\partial^{\hat{a}}\hat{\phi} \right)^{2}
+\hat{\omega}_{\hat{a}}{}^{\hat{b}\hat{c}}
\hat{\omega}_{\hat{b}\hat{c}}{}^{\hat{a}}\right]
+{\textstyle\frac{1}{4}} \left( \hat{F}^{(1)} \right)^{2}\right\}\, .
\end{eqnarray}

\noindent Finally, using Palatini's identity Eq.~(\ref{eq:Pal}) again,
but now for $d=10$ and $\phi = \hat{\phi}$, we get for the Ricci-scalar
term:

\begin{eqnarray}
\lefteqn{{\textstyle\frac{1}{2}} \int d^{11}x\
\sqrt{\hat{\hat{g}}}\ \left[ -\hat{\hat{R}} \right]=}
\nonumber \\
& &
\nonumber \\
& &
{\textstyle\frac{1}{2}} \int d^{10}x\
\sqrt{-\hat{g}}\ \left\{ e^{-2\hat{\phi}}
\left[ -\hat{R} +4\left( \partial\hat{\phi} \right)^{2}\right]
+{\textstyle\frac{1}{4}} \left( \hat{F}^{(1)} \right)^{2}\right\}\, .
\end{eqnarray}

We next reduce the $\hat{\hat{G}}$-term in Eq.~(\ref{eq:11daction}).
Usually we identify field strengths in eleven and ten dimensions with
flat indices, but in this case we also have to take into account the
scaling of the ten-di\-men\-sion\-al metric, and therefore we define

\begin{equation}
\hat{G}_{\hat{a}\hat{b}\hat{c}\hat{d}}=
e^{-\frac{4}{3}\hat{\phi}}\
\hat{\hat{G}}_{\hat{a}\hat{b}\hat{c}\hat{d}}\, ,
\end{equation}

\noindent which leads to

\begin{equation}
\hat{G} = \partial\hat{C} -2\hat{H}^{(1)}\hat{A}^{(1)}\, ,
\end{equation}

\noindent where $\hat{H}^{(1)}$ is the field strength of the
two-form $\hat{B}^{(1)}$

\begin{equation}
\hat{H}^{(1)} = \partial\hat{B}^{(1)}\, .
\end{equation}
Observe that, in spite of the fact that there is a vector field present,
the two-form field strength does not contain any Chern-Simons term.

The remaining components of ${\hat {\hat G}}$ are given by

\begin{equation}
\hat{\hat{G}}_{\hat{a}\hat{b}\hat{c} y}
= {\textstyle\frac{1}{2}}\ e^{\frac{1}{3}\hat{\phi}}
\hat{H}^{(1)}_{\hat{a}\hat{b}\hat{c}}\, ,
\end{equation}

\noindent and the contribution of the $\hat{\hat{G}}$-term to the
ten-di\-men\-sion\-al action becomes

\begin{equation}
{\textstyle\frac{1}{2}} \int d^{11} x \ \sqrt{\hat{\hat{g}}}\
a_{1} \left( \hat{\hat{G}} \right)^{2}= {\textstyle\frac{1}{2}}
\int d^{10} x\ \sqrt{-\hat{g}}\
\left[-a_{1} e^{-2\hat{\phi}}\left( \hat{H}^{(1)} \right)^{2} +a_{1}
\hat{G}^{2} \right]\, .
\end{equation}

Finally, taking into account

\begin{equation}
\hat{\hat{\epsilon}}^{\ \mu_{0}\ldots\mu_{9}{\underline y}} =
\hat{\epsilon}^{\ \mu_{0}\ldots\mu_{9}}\, ,
\end{equation}

\noindent the third term in the $d=11$ action Eq.~(\ref{eq:11daction})
(all terms with curved indices) gives

\begin{equation}
\hat{\hat{\epsilon}}\hat{\hat{G}}\hat{\hat{G}}\hat{\hat{C}}=
2 \hat{\epsilon}\partial\hat{C}\partial\hat{C}\hat{B}^{(1)}
-4\hat{\epsilon}\partial\hat{C}\partial\hat{B}^{(1)}\hat{C}\, ,
\end{equation}

\noindent and integrating by parts we get

\begin{equation}
{\textstyle\frac{1}{2}} \int d^{11} x\
a_{2} \hat{\hat{\epsilon}} \hat{\hat{G}} \hat{\hat{G}}\hat{\hat{C}} =
{\textstyle\frac{1}{2}}\int d^{10} x\
\left[ 6 a_{2}\hat{\epsilon}
\partial\hat{C}\partial\hat{C}\hat{B}^{(1)}\right]\, .
\end{equation}

Collecting all our results and setting the constants $a_{1}=3/4, a_2 =
1/384$ we find that the bosonic part of the type--IIA supergravity
action in ten dimensions in the ``string-frame'' metric is given
by\footnote{The type--IIA action in the ``Einstein-frame'' metric has
been given in \cite{kn:IIA}.}

\begin{eqnarray}
\hat{S} & = &
{\textstyle\frac{1}{2}} \int d^{10}x\
\sqrt{-\hat{g}} \left\{ e^{-2\hat{\phi}}
\left[ -\hat{R} +4\left( \partial\hat{\phi} \right)^{2}
-{\textstyle\frac{3}{4}} \left( \hat{H}^{(1)} \right)^{2}\right]\right.
\nonumber \\
& &
\nonumber \\
& &
+{\textstyle\frac{1}{4}} \left( \hat{F}^{(1)} \right)^{2}
+{\textstyle\frac{3}{4}}\hat{G}^{2}
+{\textstyle\frac{1}{64}} \frac{\hat{\epsilon}}{\sqrt{-\hat{g}}}\
\partial\hat{C}\partial\hat{C}\hat{B}^{(1)}\biggr \}\, .
\label{eq:10daction}
\end{eqnarray}

The dilaton dependence here is at first sight rather surprising.  The
first line of Eq.~(\ref{eq:10daction}) describes the fields from the
NS-NS sector and is the same as the bosonic part of the type--I
supergravity action, and in particular has the expected dilaton
dependence.  The second line of Eq.~(\ref{eq:10daction}) involves the
fields from the RR sector -- it vanishes in the truncation from
type--IIA to type--I supergravity:

\begin{equation}
\hat{C}=\hat{A}^{(1)}=0\, .
\label{eq:truncationIIA}
\end{equation}

\noindent -- and {\it is independent of the dilaton} (as noted
independently by E.~Witten \cite{kn:Wi2,kn:Wi1}).  The absence of
dilaton coupling to the RR fields reflects the fact that in {\it four
dimensions} the RR fields are invariant under $S$--duality and that the
RR charges are carried by solitons, not fundamental string modes
\cite{kn:Hu1}.

Usually, the scalar sector in supergravity theories parametrizes a
coset.  In this case, there is only one scalar and the corresponding
coset is trivially given by: $SO(1,1)/\II$ \cite{kn:third}.
Nevertheless, it leads to a global $SO(1,1)$--invariance (with parameter
$\alpha$) which describes the coupling of the dilaton.  Defining the
scale weight $w$ of a field $A$ by $A \rightarrow e^{w\alpha}A$, the
transformation rules under $SO(1,1)$ are specified by the weights in
Table~\ref{tab-weights}.  This symmetry is the $S$--duality symmetry of
the ten--dimensional type--IIA theory \cite{kn:Hu1} and so we see that
the ten-di\-men\-sion\-al RR fields do transform under $S$--duality.

\begin{table}
\begin{center}
\begin{tabular}{||l|c|l|c||}
\hline
field&weight $w$&field&weight $w$\\
\hline
& & & \\
$e^{\hat\phi}$&1&${\hat A}^{(1)}$&-3/4\\
& & & \\
$\hat g$&1/2&$\hat C$&-1/4\\
& & & \\
${\hat B}^{(1)}$&1/2&&\\
\hline
\end{tabular}
\end{center}

\caption{This table gives the weights $w$ of the fields of type--IIA
supergravity under the global $SO(1,1)$ symmetry.} \label{tab-weights}
\end{table}

It is instructive to check the gauge invariance of the action
Eq.~(\ref{eq:10daction}).  In eleven dimensions we have
reparametrization invariance and the gauge symmetry (\ref{eq:gauge11}).
{}From the ten-di\-men\-sion\-al point of view, only the the
re\-pa\-ra\-me\-tri\-za\-tion invariance of the
ele\-ven-di\-men\-sion\-al theory in the direction parametrized by the
coordinate $y$ is relevant:

\begin{equation}
\delta y =-\Lambda^{(1)}(x)\, ,
\label{eq:gauge11b}
\end{equation}

\noindent where $\Lambda^{(1)}(x)$ does not depend on $y$.  We only
consider the infinitesimal form of the gauge transformations.  Under
Eq.~(\ref{eq:gauge11b}) the ele\-ven-di\-men\-sion\-al fields transform
as follows:

\begin{equation}
\begin{array}{rclrcl}
\delta\hat{\hat{g}}_{\hat{\mu}\hat{\nu}}
&
=
&
2{\hat {\hat g}}_{\underline{y}(\hat{\mu}}
\partial_{\hat{\nu})} \Lambda^{(1)}\, ,
\hspace{.5cm}
&
\delta \hat{\hat{g}}_{\hat{\mu}\underline{y}}
&
=
&
\hat{\hat{g}}_{\underline{y}\underline{y}}
\partial_{\hat{\mu}}\Lambda^{(1)}\, ,
\\
& & & & &
\\
\delta \hat{\hat{C}}_{\hat{\mu}\hat{\nu}\hat{\rho}}
&
=
&
3\hat{\hat{C}}_{\underline{y}[\hat{\mu}\hat{\nu}}
\partial_{\hat{\rho}]}\Lambda^{(1)}\, .
& & &
\\
\end{array}
\end{equation}

\noindent These transformations and the gauge transformations of the
three-form potential Eq.~(\ref{eq:gauge11}) reduce to the following
transformations of the fields of the ten-di\-men\-sion\-al theory

\begin{eqnarray}
\delta \hat{A}^{(1)} & = & \partial\Lambda^{(1)}\, ,
\hspace {1cm}
\delta \hat{B}^{(1)}  = \partial\hat{\eta}^{(1)}\, ,
\nonumber \\
& &
\nonumber \\
\delta \hat{C} & = & \partial\hat{\chi}
+2\hat{B}^{(1)}\partial\Lambda^{(1)}\, ,
\label{eq:10dg}
\end{eqnarray}

\noindent where the parameters of the ten-di\-men\-sion\-al gauge
transformations are related to the ele\-ven-di\-men\-sion\-al parameter
$\hat{\hat{\chi}}$ by

\begin{equation}
\hat{\chi}_{\hat{\mu}\hat{\nu}} =
\hat{\hat{\chi}}_{\hat{\mu}\hat{\nu}}\, ,
\hspace{1cm}
\hat{\eta}^{(1)}_{\hat{\mu}} =
\hat{\hat{\chi}}_{\hat{\mu}\underline{y}}\, .
\end{equation}

\noindent It is easy to check that the ten-di\-men\-sion\-al action
Eq.~(\ref{eq:10daction}) is invariant under the above gauge
transformations.


\section{Reduction To Nine Dimensions}
\label{sec-10to9}

We shall now compactify the type--IIA--theory to $N=2$ supergravity in
nine dimensions to facilitate the derivation of the type--II duality
rules.  This is in accordance with the interpretation of duality as the
non-compact symmetry of a compactified supergravity theory \cite{kn:Ma1,
kn:Hu1}.

To reduce the first line of the type--IIA action given in
Eq.~(\ref{eq:10daction}) we can use the results for the heterotic string
which have been given elsewhere (see e.g.~\cite{kn:Be5}).  Since we
change notation slightly with respect to \cite{kn:Be5} we summarize some
relevant formulae here.  First we parametrize the zehnbein as follows:

\begin{equation}
\left( \hat{e}_{\hat{\mu}}{}^{\hat{a}} \right)=
\left(
\begin{array}{cc}
e_{\mu}{}^{a} & k A_{\mu}^{(2)} \\
& \\
0           & k         \\
\end{array}
\right)
\, ,
\hspace{1cm}
\left( \hat{e}_{\hat{a}}{}^{\hat{\mu}} \right)=
\left(
\begin{array}{cc}
e_{a}{}^{\mu} & -A_{a}^{(2)} \\
& \\
0           & k^{-1}   \\
\end{array}
\right)\, ,
\label{eq:basis2}
\end{equation}

\noindent where

\begin{equation}
k=\left|\hat{k}_{\hat{\mu}}\hat{k}^{\hat{\mu}}\right|^{\frac{1}{2}}\, ,
\end{equation}

\noindent and $A_{a}^{(2)}=e_{a}{}^{\mu}A_{\mu}^{(2)}$. Here ${\hat
k}_\mu$ is a Killing vector such that

\begin{equation}
{\hat k}^{{\hat \mu}}\partial_{{\hat \mu}} = \partial_{{\underline x}}\, .
\end{equation}

\noindent This time we assume that all fields are independent of the
coordinate $x = x^{\underline{9}}$ which we choose to be a space-like
direction $({\hat \eta}_{\underline {xx}}=-1$).  Note that
$\hat{k}_{\hat{\mu}}\hat{k}^{\hat{\mu}}= \hat g_{\underline {xx}} =
-k^{2}$.

Using the above zehnbeins, the ten-di\-men\-sion\-al fields $\{{\hat
g}_{\hat\mu\hat\nu}, \hat{B}^{(1)}_{\hat\mu\hat\nu}, \hat\phi\}$
decompose as follows

\begin{equation}
\begin{array}{rclrcl}
\hat{g}_{\mu\nu} & = & g_{\mu\nu}-k^{2}A_{\mu}^{(2)}A_{\nu}^{(2)}\, ,
\hspace{1cm}&
\hat{B}_{\mu\nu}^{(1)} & = & B_{\mu\nu}^{(1)}+A_{[\mu}^{(2)}B_{\nu]}\, ,
\\
& & & & &
\\
\hat{g}_{\underline{x}\underline{x}} & = & -k^{2}\, , &
\hat{B}_{\underline{x}\mu}^{(1)} & = & B_{\mu}\, ,
\\
& & & & &
\\
\hat{g}_{\underline{x}\mu} & = & -k^{2}A_{\mu}^{(2)}\, , &
\hat{\phi} & = & \phi +\frac{1}{2}\log k\, ,
\end{array}
\label{eq:DD-1}
\end{equation}

\noindent where $\left\{g_{\mu\nu},B_{\mu\nu}^{(1)},\phi,A_{\mu}^{(2)},
B_{\mu},k\right\}$ are the ni\-ne-di\-men\-sion\-al fields.  They are
given in terms of the ten-di\-men\-sion\-al fields by

\begin{equation}
\begin{array}{rclrcl}
g_{\mu\nu} & = & \hat{g}_{\mu\nu}-\hat{g}_{\underline{x}\mu}
\hat{g}_{\underline{x}\nu}/\hat{g}_{\underline{x}\underline{x}}\, ,
\hspace{1.5cm} &
B_{\mu} & = & \hat{B}_{\underline{x}\mu}^{(1)}\, ,
\\
& & & & &
\\
B_{\mu\nu}^{(1)} & = & \hat{B}_{\mu\nu}^{(1)} +
\hat{g}_{\underline{x}[\mu}\hat{B}_{\nu]\underline{x}}^{(1)}
/\hat{g}_{\underline{x}\underline{x}}\, , &
\phi & = & \hat{\phi}-\frac{1}{4}
\log{\left( -\hat{g}_{\underline{x}\underline{x}}\right)}\, ,
\\
& & & & &
\\
A_{\mu}^{(2)} & = & \hat{g}_{\underline{x}\mu}
/\hat{g}_{\underline{x}\underline{x}}\, , &
k & = & \left( -\hat{g}_{\underline{x}\underline{x}}
\right)^{\frac{1}{2}}\, .
\end{array}
\end{equation}

\noindent Therefore, ignoring the integral over $x$, the first line in
the ten-di\-men\-sion\-al action Eq.~(\ref{eq:10daction}) can be written
as

\begin{eqnarray}
{\textstyle\frac{1}{2}} \int d^{10}x\
\sqrt{-\hat{g}}\ e^{-2\hat{\phi}}\left[
 -\hat{R} +4(\partial\hat{\phi})^{2} -{\textstyle\frac{3}{4}}
\left( \hat{H}^{(1)} \right)^{2}\right] =
\hspace{-6cm}
& &
\nonumber\\
& &
\nonumber\\
& &
{\textstyle\frac{1}{2}}\int d^{9}x\  \sqrt{g}\ e^{-2\phi}
\left[-R +4(\partial\phi)^{2} -{\textstyle\frac{3}{4}}
\left( H^{(1)} \right)^{2}\right.
\\
& &
\nonumber \\
& &
\left. -(\partial\log{k})^{2} +{\textstyle\frac{1}{4}}k^{2} \left(
F^{(2)} \right)^{2} +{\textstyle\frac{1}{4}}k^{-2}F^{2}(B) \right]\, ,
\nonumber
\end{eqnarray}

\noindent where

\begin{eqnarray}
F^{(2)} & = & 2\partial A^{(2)}\, ,
\hspace{1.5cm}
F(B)  =  2\partial B\, ,
\nonumber \\
& &
\\
H^{(1)} & = & \partial B^{(1)}  +
A^{(2)}\partial B + B \partial A^{(2)}\, .
\nonumber
\end{eqnarray}

We next reduce the first term in the second line of
Eq.~(\ref{eq:10daction}).  The vector field ${\hat A}^{(1)}$ reduces to
a scalar and a vector as follows:

\begin{eqnarray}
{\hat A}^{(1)}_{{\underline x}} & = & \ell\, ,
\nonumber \\
& &
\\
{\hat A}^{(1)}_\mu & = & A^{(1)}_\mu + \ell A_\mu^{(2)}\, .
\nonumber
\end{eqnarray}

\noindent We thus find:

\begin{eqnarray}
\int d^{10}x\
\sqrt{-\hat{g}}\ \left[ {\textstyle\frac{1}{2}}
\left( {\hat F}^{(1)} \right)^{2} \right]=
\hspace{-4cm}
\nonumber \\
& &
\\
& &
\int d^{9}x\ \sqrt{g}\ \left[
{\textstyle\frac{1}{4}} k \left( F^{(1)} + \ell F^{(2)} \right)^{2}
-{\textstyle\frac{1}{2}}k^{-1}(\partial \ell)^{2} \right]\, .
\nonumber
\end{eqnarray}

\noindent To reduce the $\hat{G}^{2}$ term in Eq.~(\ref{eq:10daction})
we decompose the three--index tensor $\hat{C}$ as follows:

\begin{eqnarray}
\hat{C}_{\mu\nu\underline{x}} & = & {\textstyle\frac{2}{3}} \left(
B_{\mu\nu}^{(2)} - A_{[\mu}^{(1)}B_{\nu]}\right)
\nonumber\\
& &
\\
\hat{C}_{\mu\nu\rho} & = & C_{\mu\nu\rho}\, .
\nonumber
\end{eqnarray}

\noindent For the sake of completeness we also give the expression of
the ni\-ne-di\-men\-sion\-al fields
$C_{\mu\nu\rho},B^{(2)}_{\mu\nu},A^{(1)}_{\mu},\ell$ in terms of the
ten-di\-men\-sion\-al ones:

\begin{equation}
\begin{array}{rclrcl}
C_{\mu\nu\rho}
&
=
&
\hat{C}_{\mu\nu\rho}\, ,\hspace{1cm}
&
B^{(2)}_{\mu\nu}
&
=
&
\frac{3}{2}\hat{C}_{\mu\nu\underline{x}}
-\hat{A}^{(1)}_{[\mu}\hat{B}^{(1)}_{\nu]\underline{x}}
+\hat{g}_{\underline{x}[\mu}\hat{B}^{(1)}_{\nu]\underline{x}}
\hat{A}^{(1)}_{\underline{x}}/\hat{g}_{\underline{x}\underline{x}}\, ,\\
& & & & & \\
\ell
&
=
&
\hat{A}^{(1)}_{\underline{x}}\, ,
&
A_{\mu}^{(1)}
&
=
&
\hat{A}^{(1)}_{\mu} -\hat{A}^{(1)}_{\underline{x}}
\hat{g}_{\underline{x}\mu}/\hat{g}_{\underline{x}\underline{x}}\, .
\\
\end{array}
\end{equation}
We find that

\begin{equation}
\hat{G}_{abcx} = {\textstyle\frac{1}{2}} k^{-1}\left(
H_{abc}^{(2)} - \ell H_{abc}^{(1)} \right)\, ,
\end{equation}

\noindent with

\begin{equation}
H^{(2)}  = \partial B^{(2)} - A^{(1)}\partial B - B\partial A^{(1)}\, .
\end{equation}

\noindent At this point it is convenient to make use of the global
$O(2)$--invariance of the $N=2, d=9$ supergravity theory explained in
Section~\ref{sec-dualityin10} (see also Appendix~\ref{sec-11vs9}) and to
write the field-strengths $H^{(1)}$ and $H^{(2)}$ as

\begin{equation}
H^{(i)} = \partial B^{(i)} + \epsilon^{ij}\left(
A^{(j)}\partial B + B\partial A^{(j)}\right)\, , \hskip 1truecm
i=1,2,\,\, \epsilon^{12}=-\epsilon^{21}=+1\, .
\label{eq:9dH}
\end{equation}

\noindent Similarly we find that

\begin{equation}
{\hat G}_{abcd} = G_{abcd}\, ,
\end{equation}

\noindent with

\begin{equation}
G = \partial C +2 A^{(i)}\partial B^{(i)} -2\epsilon^{ij}BA^{(i)}
\partial A^{(j)}\, .
\end{equation}

\noindent We thus find that the $\hat{G}^{2}$ term in
Eq.~(\ref{eq:10daction}) is given by

\begin{equation}
\int d^{10}x\
\sqrt{-\hat{g}}\ \left[ {\textstyle\frac{3}{4}} \hat{G}^{2} \right] =
\int d^{9} x\sqrt{g}\ \left[
{\textstyle\frac{3}{4}} k G^{2}-{\textstyle\frac{3}{4}}k^{-1}
\left( H^{(2)} -\ell H^{(1)}\right)^{2}\right]\, .
\end{equation}

\noindent Finally, we reduce the $\hat{\epsilon}\partial\hat{C}\partial
\hat {C} \hat{B}^{(1)}$ term in Eq.~(\ref{eq:10daction}).  A
straightforward application of the previous formulae gives:

\begin{eqnarray}
\hat{\epsilon}\partial\hat{C}\partial \hat{C}\hat{B}^{(1)}
& = &
\epsilon\  \left\{ -2\partial C\partial C B -2\partial
C \partial B^{(i)} B^{(j)} \epsilon^{ij} \right.
\nonumber\\
& &
\nonumber\\
& &
\left. -4\partial C A^{(i)}\partial B^{(i)}B
+2\partial C A^{(i)} A^{(j)}\partial B B \epsilon^{ij}\right\}\, .
\end{eqnarray}

In summary, the fields of the $N=2, d=9$ supergravity theory are given
by:

\begin{equation}
\left\{g_{\mu\nu}, C_{\mu\nu\rho}, B^{(i)}_{\mu\nu}, A^{(i)}_\mu,
B_\mu, \phi, k, \ell\right\}
\end{equation}

\noindent The action for these fields is given by

\begin{eqnarray}
S  & = &
{\textstyle\frac{1}{2}}\int d^{9}x\  \sqrt{g}\ \left\{ e^{-2\phi}
\left[ -R +4(\partial\phi)^{2} -{\textstyle\frac{3}{4}}
\left( H^{(1)} \right)^{2} \right. \right.
\nonumber \\
& &
\nonumber \\
& &
\left. -(\partial\log k)^{2} +{\textstyle\frac{1}{4}}k^{2}\left( F^{(2)}
\right)^{2} +{\textstyle\frac{1}{4}} k^{-2} F^{2}(B)\right]
\nonumber \\
& &
\nonumber \\
& &
+{\textstyle\frac{1}{4}} k \left( F^{(1)} + \ell F^{(2)} \right)^{2}
-{\textstyle\frac{1}{2} }k^{-1} (\partial \ell)^{2}
\nonumber\\
& &
\nonumber\\
& &
+{\textstyle\frac{3}{4}} k G^{2}
-{\textstyle{3}{4}} k^{-1} \left( H^{(2)} -\ell H^{(1)}\right)^{2}
\nonumber \\
& &
\nonumber \\
& &
-{\textstyle\frac{1}{32}}\frac{1}{\sqrt{g}}\ \epsilon\
\left( \partial C\partial C B + \partial C\partial B^{(i)}
B^{(j)}\epsilon^{ij}\right.
\nonumber\\
& &
\nonumber\\
& &
\left.\left.+2\partial C A^{(i)}\partial B^{(i)}B
-\partial C A^{(i)} A^{(j)}\partial B B\epsilon^{ij}\right)\right\}\, .
\label{eq:9daction}
\end{eqnarray}

In Ref.~\cite{kn:Cr1}, it was suggested that the $N=2,d=9$ supergravity
action should have a global $GL(2,\R)=SL(2,\R)\times SO(1,1)$
invariance\footnote{Any matrix of the group $GL(2,\R)$ can be uniquely
written as the product of an $SL(2,\R)$ matrix, a real positive number
and $+1$ or $-1$.  This gives the decomposition $GL(2,\R)=SL(2,\R)\times
\R^{+}\times\Z_{2}$.  (The multiplicative group of the real positive
number $\R^{+}$ is isomorphic to the aditive group $\R$.)  Finally,
$SO(1,1)=\R^{+}\times\Z_{2}$.}.  However, oh physical grounds, one would
expect a symmetry group containing at least the $S$--duality group
$SL(2,\R)$ and the $T$--duality group $O(1,1)=SO(1,1)\times\Z_{2}$, that
is, $GL(2,\R)\times\Z_{2}$.  As we shall see in
Section~\ref{sec-dualityin10}, the invariance is indeed $GL(2,\R)$ and
the \lq missing' $\Z_2$ invariance will be the main theme of
Section~\ref{sec-dualityin10}; it is related to the $T$--duality of the
type--II theory.

It is instructive to consider the gauge invariances of this action.  In
ten dimensions we have the reparametrizations in the $x$-direction with
a parameter $\Lambda^{(2)}(x^{\mu})$ independent of $x$ and the gauge
transformations Eq.~(\ref{eq:10dg}).  After dimensional reduction they
become the following symmetries of the ni\-ne-di\-men\-sion\-al theory:

\begin{eqnarray}
\delta A^{(i)} & = & \partial \Lambda^{(i)}\, ,
\hspace {1cm}
\delta B  =  \partial \Lambda\, ,
\nonumber \\
& &
\nonumber \\
\delta B^{(i)} & = & \partial \eta^{(i)} - \epsilon^{ij}
\left( B\partial\Lambda^{(j)} + A^{(j)}\partial\Lambda\right)\, ,
\\
& &
\nonumber \\
\delta C & = & \partial\chi + 2 B^{(i)}\partial \Lambda^{(i)}
+ 2B A^{(i)}\partial\Lambda^{(j)}\epsilon^{ij}\, ,
\nonumber
\end{eqnarray}

\noindent where the parameters of the ni\-ne-di\-men\-sion\-al gauge
transformations are related to the ten-di\-men\-sion\-al parameters by

\begin{eqnarray}
\Lambda & = & -{\textstyle\frac{1}{2}}
\hat{\eta}^{(1)}_{\underline{x}}\, , \hspace{1cm}
\chi_{\mu\nu}  =  \hat{\chi}_{\mu\nu}\, ,
\nonumber \\
& &
\nonumber \\
\eta_{\mu}^{(1)} & = & \hat{\eta}_{\mu}^{(1)}\, ,
\hspace {1cm}
\eta_{\mu}^{(2)}  =  \hat{\chi}_{\mu\underline{x}}\, .
\end{eqnarray}
It is straightforward to check that the ni\-ne-di\-men\-sion\-al action
Eq.~(\ref{eq:9daction}) is invariant under the above gauge
transformations.


\section{The Type--IIB Superstring}
\label{sec-IIB}

We shall also need to consider the low-energy limit of the type--IIB
superstring for our discussion duality.  The zero-slope limit of the
type--IIB superstring is given by $N=2, d=10$ chiral supergravity
\cite{kn:IIB1,kn:IIB2}.  This theory contains a metric, a complex
antisymmetric tensor, a complex scalar and a four-index antisymmetric
tensor gauge field.  The complex scalar parametrizes the coset
$SU(1,1)/U(1)$.  In order to distinguish between the type--IIA and
type--IIB fields, we denote the type--IIB fields as follows:

\begin{equation}
\left\{
\hat{D}_{\hat{\mu}\hat{\nu}\hat{\lambda}\hat{\rho}},
\hat{h}_{\hat{\mu}\hat{\nu}},
\hat{{\cal B}}_{\hat{\mu}\hat{\nu}},\hat{\Phi}
\right\}\, ,
\end{equation}

\noindent where $\hat{h}_{\hat{\mu}\hat{\nu}}$ is the ``Einstein-frame''
metric.  We will start in the ``Einstein-frame'' and then switch to the
``string-frame'' metric once we have correctly identified the type--IIB
dilaton field.

The field equations of the type--IIB theory cannot be derived from a
covariant action.  The type--IIB field equations of Ref.~\cite{kn:IIB1}
are given (in our notation and conventions) by

\begin{eqnarray}
\hat{R}_{\hat{\mu}\hat{\nu}}\left( \hat{h} \right)
& = &
-2\hat{P}_{(\hat{\mu}} \hat{P}^{*}_{\hat{\nu})}
-{\textstyle\frac{25}{6}}
\hat{F} \left( \hat{D} \right)_{\hat{\lambda}_{1}\cdots
\hat{\lambda}_{4}\hat{\mu}}
\hat{F} \left( \hat{D} \right)^{\hat{\lambda}_{1}\cdots
\hat{\lambda}_{4}}{}_{\hat{\nu}}
\nonumber \\
& &
\nonumber \\
& &
-{\textstyle\frac{9}{4}}
\hat{G}_{(\hat{\mu}}{}^{\hat{\lambda}\hat{\rho}}
\hat{G}^{*}_{\hat{\nu}) \hat{\lambda}\hat{\rho}}
+{\textstyle\frac{3}{16}} \hat{h}_{\hat{\mu}\hat{\nu}}
\hat{G}\hat{G}^{*}\, ,
\nonumber\\
& &
\nonumber\\
\nabla^{\hat{\lambda}}\hat{G}_{\hat{\mu}\hat{\nu}\hat{\lambda}}
& = &
{\textstyle\frac{1}{2}} \hat{Q}^{\hat{\lambda}}
\hat{G}_{\hat{\mu}\hat{\nu}\hat{\lambda}}
+\hat{P}^{\hat{\lambda}}\hat{G}^{*}_{\hat{\mu}\hat{\nu}\hat{\lambda}}
-{\textstyle\frac{10}{3}} i \hat{F} \left( \hat{D}
\right)_{\hat{\mu}\hat{\nu}\hat{\lambda}\hat{\rho}\hat{\sigma}}
\hat{G}^{\hat{\lambda}\hat{\rho}\hat{\sigma}}\, ,
\nonumber\\
& &
\nonumber\\
\nabla^{\hat{\mu}}\hat{P}_{\hat{\mu}}
& = &
\hat{Q}^{\hat{\lambda}} \hat{P}_{\hat{\lambda}}
-{\textstyle\frac{3}{8}} \hat{G}^{2}\, ,
\\
& &
\nonumber \\
\hat{F} \left(\hat{D}\right)
& = &
\tilde{\hat{F}} \left(\hat{D}\right)\, .
\nonumber
\end{eqnarray}

\noindent We have used here the following definitions:

\begin{eqnarray}
\hat{G}
& = &
\frac{\hat{\cal H} -\hat{\Phi} \hat{\cal H}^{*}}{\left(
1-\hat{\Phi}^{*}\hat{\Phi} \right)^{1/2}}\, ,
\hskip .5truecm {\rm with}
\hskip .5truecm {\cal H}=\partial {\cal B}\, ,
\nonumber\\
& &
\nonumber\\
\hat{F} \left( \hat{D} \right)
& = &
\partial\hat{D} -{\textstyle\frac{3}{8i}}
\left( \hat{\cal B} \partial \hat{\cal B}^{*}
-\hat{\cal B}^{*}\partial \hat{\cal B}\right)\, ,
\\
& &
\nonumber \\
\hat{P}
& = &
\frac{\partial\hat{\Phi}}{1-\hat{\Phi}^{*}\hat{\Phi}}\, ,
\hskip 1truecm
\hat{Q} = \frac{\hat{\Phi} \overset{\leftrightarrow}{\partial}
\hat{\Phi}^{*}}{1-\hat{\Phi}^{*}\hat{\Phi}}\, .
\nonumber
\end{eqnarray}

\noindent The theory is invariant under $d=10$ general coordinate
transformations and under the following tensor gauge transformations:

\begin{eqnarray}
\delta \hat{\cal B} & = & \partial \hat{\Sigma}\, ,
\nonumber\\
& &
\nonumber\\
\delta \hat{D} & = & \partial\hat{\rho} +{\textstyle\frac{3}{8i}}
\left( \partial\hat{\Sigma}\hat{\cal B}^{*}
-\partial\hat{\Sigma}^{*}\hat{\cal B} \right)\, .
\end{eqnarray}

It is known that the dimensional reduction of $d=10$ type--IIA and IIB
supergravity leads to the same $N=2, d=9$ supergravity theory.  Our task
is to make the correct identifications between the dimensionally reduced
type--IIB fields and the fields of $N=2, d=9$ supergravity as found in
the previous section.  It is convenient to start by rewriting the theory
using the ``string-frame'' metric $\hat{\jmath}_{\hat{\mu}\hat{\nu}}$,
but before we have to identify the type--IIB dilaton.  This is easier to
do in the $SL(2,\R)$ version of the theory.  Accordingly, we first
define the complex scalar field
$\hat{\lambda}=\hat{\ell}+ie^{-\hat{\varphi}}$ by

\begin{equation}
-i\hat{\lambda}= \frac{1-\hat{\Phi}}{1+\hat{\Phi}}\, ,
\end{equation}

\noindent which gives

\begin{equation}
\frac{\partial_{\hat{\mu}}\hat{\Phi}
\partial^{\hat{\mu}}\hat{\Phi}^{*}}{(1
-\hat{\Phi}\hat{\Phi}^{*})^{2}}
=\frac{1}{4} \frac{\partial_{\hat{\mu}} \hat{\lambda}
\partial^{\hat{\mu}} \hat{\lambda}^{*}}{(\Im{\rm m}\hat{\lambda})^{2}}\,
, \end{equation}

\noindent so $\hat{\lambda}$ parametrizes an $SL(2,\R)$ coset.  We next
define the ``string-frame'' metric $\hat{\jmath}_{\hat{\mu}\hat{\nu}}$
by

\begin{equation}
\hat{\jmath}_{\hat{\mu}\hat{\nu}}=e^{\frac{1}{2}\hat{\varphi}}
\hat{h}_{\hat{\mu}\hat{\nu}}\, .
\end{equation}

\noindent This definition implies that $\hat{\varphi}$ is the type--IIB
dilaton and will be justified below.  We next consider the complex
antisymmetric tensor ${\cal B}$.  To make contact with the ``real''
$O(2)$ notation of the previous section we write

\begin{equation}
\hat{\cal B} = \hat{\cal B}^{(1)} +i\hat{\cal B}^{(2)}\, ,
\hskip 1truecm
\hat{\Sigma} = \hat{\Sigma}^{(1)} +i\hat{\Sigma}^{(2)}\, .
\end{equation}

\noindent Using this notation the field-strengths of the $\hat {\cal B}$
gauge fields and their gauge transformations can be written as:

\begin{equation}
\begin{array}{rclrcl}
\hat{\cal H}^{(i)}
&
=
&
\partial\hat{\cal B}^{(i)}\, ,
&
\delta\hat{\cal B}^{(i)}
&
=
&
\partial\hat{\Sigma}^{(i)}\, ,
\\
& & & & &
\\
\hat{F}\left( \hat{D} \right)
&
=
&
\partial\hat{D} +{\textstyle\frac{3}{4}}
\epsilon^{ij}\hat{\cal B}^{(i)} \partial\hat{\cal B}^{(j)}\, ,
&
\delta\hat{D}
&
=
&
\partial\hat{\rho} -{\textstyle\frac{3}{4}}
\epsilon^{ij}\partial\hat{\Sigma}^{(i)} \hat{\cal B}^{(j)}\, .
\end{array}
\end{equation}

To explain why it is appropriate to identify the type--IIB dilaton with
the $\hat{\varphi}$ scalar field it is convenient to use the following
trick.  Although there is no action in ten dimensions giving rise to the
full type--IIB field equations it turns out that one can write down an
action giving rise to the type--IIB field equations with
$\hat{F}(\hat{D}) = 0$.  This action is given by:

\begin{eqnarray}
\hat{S}_{IIB}^{{\rm sugra}} & = &
{\textstyle\frac{1}{2}} \int d^{10}x\
\sqrt{-\hat{h}}\ \left[ -\hat{R} \left( \hat{h} \right)
-2\frac{\partial\hat{\Phi}\partial\hat{\Phi}^{*}}{
\left( 1-\hat{\Phi}^{*}\hat{\Phi} \right)^{2}}
-{\textstyle\frac{3}{4}} \hat{G}^{*} \hat{G}\right]\, .
\label{eq:IIBsugra}
\end{eqnarray}

\noindent If we now perform all the above changes in this action we get
the following action in the ``string-frame'' metric:

\begin{eqnarray}
\hat{S}_{IIB}^{{\rm string}} & = &
{\textstyle\frac{1}{2}} \int d^{10}x
\sqrt{-\hat{\jmath}}
\left\{ e^{-2\hat{\varphi}}
\left[ -\hat{R} \left( \hat{\jmath} \right)
+4(\partial\hat{\varphi})^{2} -{\textstyle\frac{3}{4}}
\left(\hat{\cal H}^{(1)} \right)^{2} \right] \right.
\nonumber \\
& &
\nonumber \\
& & \left. -{\textstyle\frac{1}{2}} (\partial\hat{\ell})^{2}
 -{\textstyle\frac{3}{4}} \left( \hat{\cal H}^{(2)} -\hat{\ell}
\hat{\cal H}^{(1)} \right)^{2} \right\}\, .
\label{eq:IIBstring}
\end{eqnarray}

\noindent It is easy to read from this action that the truncation
$\hat{D}=\hat{\cal B}^{(2)}=\hat{\ell}=0$ (which implies
$\hat{F}(\hat{D})=0$, so it is consistent to use this action) gives the
usual type--I action.  We see that, as in the type--IIA case, the
type--IIB RR fields do not appear multiplied by the string coupling
constant (the dilaton).

The equations of motion for the full type--IIB theory written in terms
of the ``stringy" fields

\begin{equation}
\left\{\hat{D}_{\hat{\mu}\hat{\nu}\hat{\rho}\hat{\sigma}},
\hat{\jmath}_{\hat{\mu}\hat{\nu}},
\hat{\cal B}^{(i)}_{\hat{\mu}\hat{\nu}},
\hat{\ell},\hat{\varphi}\right\}
\end{equation}

\noindent are

\begin{eqnarray}
\hat{R}_{\hat{\mu}\hat{\nu}}(\hat{\jmath})
& = &
2\nabla_{\hat{\mu}}\partial_{\hat{\nu}}\hat{\varphi}
-{\textstyle\frac{9}{4}}
\hat{\cal H}^{(1)}_{(\hat{\mu}}{}^{\hat{\lambda}\hat{\rho}}
\hat{\cal H}^{(1)}_{\hat{\nu})\hat{\lambda}\hat{\rho}}
-e^{2\hat{\varphi}}\left\{
{\textstyle\frac{1}{2}}\left[
\partial_{\hat{\mu}}\hat{\ell}\partial_{\hat{\nu}}\hat{\ell}
-{\textstyle\frac{1}{2}}\hat{\jmath}_{\hat{\mu}\hat{\nu}}(\partial
\hat{\ell})^{2}\right]\right.
\nonumber \\
& &
\nonumber \\
& &
\hspace{-1cm}
+{\textstyle\frac{9}{4}}\left[
\left( \hat{\cal H}^{(2)} -\hat{\ell}\hat{\cal H}^{(1)}
\right)_{(\hat{\mu}}{}^{\hat{\lambda}\hat{\rho}}
\left( \hat{\cal H}^{(2)} -\hat{\ell}\hat{\cal H}^{(1)}
\right)_{\hat{\nu})\hat{\lambda}\hat{\rho}}
-{\textstyle\frac{1}{6}}
\hat{\jmath}_{\hat{\mu}\hat{\nu}}\left( \hat{\cal H}^{(2)}
-\hat{\ell}\hat{\cal H}^{(1)}\right)^{2}\right]
\nonumber \\
& &
\nonumber \\
& &
\hspace{-1cm}
\left.
+{\textstyle\frac{25}{6}}
\hat{F} \left( \hat{D} \right)_{\hat{\lambda}_{1}\cdots
\hat{\lambda}_{4}\hat{\mu}}
\hat{F} \left( \hat{D} \right)^{\hat{\lambda}_{1}\cdots
\hat{\lambda}_{4}}{}_{\hat{\nu}}
\right\}\, ,
\nonumber\\
& &
\nonumber\\
\nabla^{2}\hat{\varphi}
& = &
{\textstyle\frac{1}{4}} \hat{R}(\hat{\jmath})
+{\textstyle\frac{3}{16}}\left( \hat{\cal H}^{(1)}\right)^{2}
+(\partial\hat{\varphi})^{2}\, ,
\nonumber\\
& &
\nonumber\\
\nabla^{2}\hat{\ell}
& = &
-{\textstyle\frac{3}{2}}\hat{\cal H}^{(1)}
\left(\hat{\cal H}^{(2)} -\hat{\ell}\hat{\cal H}^{(1)}
\right)\, ,
\nonumber \\
& &
\nonumber \\
\nabla^{\hat{\mu}}
\left[\left(\hat{\ell}^{2} +e^{-2\hat{\varphi}}\right)
\hat{\cal H}^{(1)} -\hat{\ell}\hat{\cal H}^{(2)}
\right]_{\hat{\mu}\hat{\nu}\hat{\rho}}
=
{\textstyle\frac{10}{3}}
\hat{F}\left(\hat{D}
\right)_{\hat{\nu}\hat{\rho}\hat{\sigma}\hat{\lambda}\hat{\kappa}}
\hat{\cal H}^{(2)\hat{\sigma}\hat{\lambda}\hat{\kappa}}\, ,
\hspace{-9cm}
& &
\nonumber \\
& &
\nonumber \\
\nabla^{\hat{\mu}}
\left(\hat{\cal H}^{(2)} -\hat{\ell}\hat{\cal H}^{(1)}
\right)_{\hat{\mu}\hat{\nu}\hat{\rho}}
=
-{\textstyle\frac{10}{3}}
\hat{F}\left(\hat{D}
\right)_{\hat{\nu}\hat{\rho}\hat{\sigma}\hat{\lambda}\hat{\kappa}}
\hat{\cal H}^{(1)\hat{\sigma}\hat{\lambda}\hat{\kappa}}\, ,
\hspace{-7cm}
& &
\nonumber \\
& &
\nonumber \\
\hat{F} \left( \hat{D} \right)
& = &
\tilde{\hat{F}} \left( \hat{D}\right)\, .
\label{eq:stringyIIB}
\end{eqnarray}

In the second equation of Eqs.~(\ref{eq:stringyIIB}) we can see that,
although the RR fields do not couple directly to the dilaton, they
couple indirectly to it through the metric.

This is going to be our starting point for the dimensional reduction to
$d=9$.  First we want the dimensional reduction of $\hat{\cal H}^{(i)}$
to reproduce the nine-\-di\-men\-sio\-nal field-strengths $H^{(i)}$
given in Eq.~(\ref{eq:9dH}).  We observe that $\hat{\cal H}^{(i)}$
contains no Chern-Simons term while $H^{(i)}$ does.  This means that in
the type--IIB reduction one of the vector fields present in the
Chern-Simons part of $H^{(i)}$ must be identified with the vector field
present in the parametrization of the type--IIB zehnbein.  In the
type--IIA reduction this vector field was called $A^{(2)}$ (see
Eq.~(\ref{eq:basis2})).  Note that the vector field $A^{(2)}$ is present
in $H^{(2)}$ but not in $H^{(1)}$ so we cannot use the same
parametrization\footnote{The situation in the type--IIA reduction is
different since there $B^{(2)}$ is related to $\hat{C}$ whose
field-strength already contains a Chern-Simons term in ten dimensions.}.
We see that on the other hand the vector field $B$ does occur in the
Chern-Simons part of both $H^{(1)}$ and $H^{(2)}$.  Therefore $B$ must
occur in the parametrization of the type--IIB zehnbein.  At this point
we realize that the NS-NS string part of the ni\-ne-di\-men\-sion\-al
action (i.e.~the first two lines in Eq.~(\ref{eq:9daction}) are
invariant under the $\Z_{2}$ transformation

\begin{equation}
\tilde{A}^{(2)}_{\mu} = {B}_\mu\, ,\hskip 1truecm
\tilde{B}_{\mu} = A^{(2)}_{\mu}\, ,\hskip 1truecm \tilde{k} = k^{-1}\, .
\end{equation}

This means that a ``dual'' parametrization of the zehnbein with
$A^{(2)}$ replaced by $B$ and $k$ replaced by $k^{-1}$ leads to the same
NS-NS part of the action Eq.~(\ref{eq:9daction}).  We therefore take the
parametrization of the ``string-frame'' type--IIB zehnbein
$\hat{\epsilon}_{\hat{\mu}}{}^{\hat{a}}$

\begin{equation}
\hat{\epsilon}_{\hat{\mu}}{}^{\hat{a}}
\hat{\epsilon}_{\hat{\nu}}{}^{\hat{b}}\hat{\eta}_{\hat{a}\hat{b}}
=\hat{\jmath}_{\hat{\mu}\hat{\nu}}\, ,
\hspace{1.5cm}
\hat{\epsilon}_{\hat{a}}{}^{\hat{\mu}}
\hat{\epsilon}_{\hat{b}}{}^{\hat{\nu}}
\hat{\jmath}^{\hat{\mu}\hat{\nu}} = \hat{\eta}_{\hat{a}\hat{b}}\, ,
\end{equation}

\noindent to be:

\begin{equation}
\left( \hat{\epsilon}_{\hat{\mu}}{}^{\hat{a}} \right)=
\left(
\begin{array}{cc}
e_{\mu}{}^{a} & k^{-1} B_{\mu} \\
& \\
0           & k^{-1}         \\
\end{array}
\right)
\, ,
\hspace{1cm}
\left( \hat{\epsilon}_{\hat{a}}{}^{\hat{\mu}}\right)=
\left(
\begin{array}{cc}
e_{a}{}^{\mu} & -B_{a} \\
& \\
0           & k   \\
\end{array}
\right)\, .
\label{eq:basis3}
\end{equation}

\noindent The gauge field $B$ transforms as $\delta B = \partial
\Lambda$ provided that we identify $\xi^{\underline{x}} = \Lambda$.

Using the parametrization Eq.~(\ref{eq:basis3}), it is a straightforward
exercise to verify that the ten-di\-men\-sion\-al gauge-invariant
fields--strengths $\hat{\cal H}^{(i)}$ decompose into the
ni\-ne-di\-men\-sion\-al gauge-invariant field-strengths $H^{(i)}$ and
$F^{(i)}$ defined in the previous section, provided that we make the
following identifications

\begin{equation}
\begin{array}{rclrcl}
\hat{\cal B}_{\mu\nu}^{(i)}
&
=
&
B_{\mu\nu}^{(i)} + \epsilon^{ij}B_{[\mu}A_{\nu]}^{(j)}\, ,
\hspace{1cm}
&
\hat{\Sigma}^{(i)}_{\mu}
&
=
&
\eta_{\mu}^{(i)}\, ,
\\
& & & & &
\\
\hat{\cal B}^{(i)}_{\underline{x}\mu}
&
=
&
\epsilon^{ij}A_{\mu}^{(j)}\, ,
&
\hat{\Sigma}^{(i)}_{\underline{x}}
&
=
&
-2\epsilon^{ij}\Lambda^{(j)}\, .
\end{array}
\end{equation}

\noindent Similarly, one may verify that type--IIB gauge field $\hat{D}$
reduces to the ni\-ne-di\-men\-sion\-al gauge field $C$ with the same
gauge transformation properties provided that we identify:

\begin{equation}
\hat{D}_{\mu\nu\rho\underline{x}} =
{\textstyle\frac{3}{8}}\left(
C_{\mu\nu\rho} - A_{[\mu}^{(i)}B_{\nu\rho]}^{(i)}-\epsilon^{ij}
A_{[\mu}^{(i)}A_{\nu}^{(j)}B_{\rho]}\right)\, ,
\hspace{.5cm}
\hat{\rho}_{\mu\nu\underline{x}} =
{\textstyle\frac{1}{2}}\chi_{\mu\nu}\, .
\end{equation}

Observe that $\hat{D}_{\mu\nu\rho\sigma}$ is not an independent
ni\-ne-di\-men\-sion\-al field. It is completely determined by
$\hat{D}_{\underline{x}\nu\rho\sigma}$ and the other fields and
therefore we will consistently ignore it from now on.

We conclude this section by giving all the relations between the
ten-di\-men\-sion\-al ``string-frame" type--IIB supergravity fields and
the nine--di\-men\-sio\-nal ones

\begin{equation}
\begin{array}{rclrcl}
\hat{D}_{\mu\nu\rho\underline{x}}
&
=
&
\frac{3}{8}\left(
C_{\mu\nu\rho} - A_{[\mu}^{(i)}B_{\nu\rho]}^{(i)}-\epsilon^{ij}
A_{[\mu}^{(i)}A_{\nu}^{(j)}B_{\rho]}\right)\, ,
\hspace{-2cm}
& & & \\
& & & & & \\
\hat{\jmath}_{\mu\nu}
&
=
&
g_{\mu\nu} -k^{-2}B_{\mu}B_{\nu}\, ,\hspace{1.5cm}
&
\hat{\jmath}_{\underline{x}\mu}
&
=
&
-k^{-2}B_{\mu}\, ,
\\
& & & & & \\
\hat{\cal B}^{(i)}_{\mu\nu}
&
=
&
B^{(i)}_{\mu\nu} +\epsilon^{ij}B_{[\mu}A_{\nu]}^{(j)}\, ,
&
\hat{\cal B}^{(i)}_{\underline{x}\mu}
&
=
&
\epsilon^{ij}A_{\mu}^{(j)}\, ,
\\
& & & & & \\
\hat{\jmath}_{\underline{x}\underline{x}}
&
=
&
-k^{-2}\, ,
&
\hat{\ell}
&
=
&
\ell\, ,
\\
& & & & & \\
\hat{\varphi}
&
=
&
\phi -\frac{1}{2}\log{k}\, ,
& & & \\
\end{array}
\end{equation}

\noindent and vice versa

\begin{equation}
\begin{array}{rclrcl}
C_{\mu\nu\rho}
&
=
&
\frac{8}{3}\hat{D}_{\underline{x}\mu\nu\rho}
+\epsilon^{ij}\hat{\cal B}^{(i)}_{\underline{x} [\mu}
\hat{\cal B}^{(j)}_{\nu\rho]}
+2\epsilon^{ij} \hat{\cal B}^{(i)}_{\underline{x} [\mu}
\hat{\cal B}^{(j)}_{|\underline{x} |\nu}
\hat{\jmath}_{\rho]\underline{x}}
/\hat{\jmath}_{\underline{x}\underline{x}}\, ,
\hspace{-5cm}
& & & \\
& & & & & \\
g_{\mu\nu}
&
=
&
\hat{\jmath}_{\mu\nu} -\hat{\jmath}_{\underline{x}\mu}
\hat{\jmath}_{\underline{x}\nu}
/\hat{\jmath}_{\underline{x}\underline{x}}\, ,
\hspace{2cm}
&
B^{(i)}_{\mu\nu}
&
=
&
\hat{\cal B}^{(i)} +\hat{\jmath}_{\underline{x} [\mu}
\hat{\cal B}^{(i)}_{\nu]\underline{x}}
/\hat{\jmath}_{\underline{x}\underline{x}}\, ,
\\
& & & & & \\
B_{\mu}
&
=
&
\hat{\jmath}_{\underline{x}\mu}
/\hat{\jmath}_{\underline{x}\underline{x}}\, ,
&
A^{(i)}
&
=
&
-\epsilon^{ij}\hat{\cal B}^{(j)}_{\underline{x}\mu}\, ,
\\
& & & & & \\
k
&
=
&
\left(-\hat{\jmath}_{\underline{x}\underline{x}}
\right)^{-\frac{1}{2}}\, ,
&
\ell
&
=
&
\hat{\ell}\, ,
\\
& & & & & \\
\phi
&
=
&
\hat{\varphi} -\frac{1}{4}
\log{(-\hat{\jmath}_{\underline{x}\underline{x}})}\, .
& & & \\
\end{array}
\end{equation}


\section{Type--II $S$-- and $T$--duality}
\label{sec-dualityin10}

In this section we shall find the type--II $S$-- and $T$--duality rules
described in the introduction.  We start by exploring the non-compact
symmetries of the type--II supergravity theory in nine dimensions and
then seek their analogues in the ``parent'' theories in ten and (in the
next section) eleven dimensions.

We start by considering the $SL(2,\R)$ $S$--duality symmetry.  The
$SL(2,\R)$ symmetry of the type--IIB theory in $d=10$ gives rise to an
$SL(2,\R)$ symmetry of the $N=2$ theory in $d=9$.  An $O(2)$ subgroup of
this is a manifest symmetry of the action (\ref{eq:9daction}).  Under
$SL(2,\R)$, $A^{(i)}_\mu$ and $B^{(i)}_{\mu \nu }$ are both doublets
while $ {\lambda}={\ell}+ie^{-{\varphi}}$ is a complex coordinate on
$SL(2,\R)/ U(1)$ transforming by fractional linear transformations.  The
origin of this $SL(2,\R)$ symmetry from the type--IIA theory is more
subtle.  An $SO(1,1)$ subgroup which acts by shifting the dilaton arises
from the $SO(1,1)$ symmetry of the type--IIA theory in $d=10$ discussed
in Section~\ref{sec-11to10}.  An $O(2)$ subgroup has a natural
interpretation as Lorentz transformations of the {\it eleven
dimensional} supergravity in a background with two commuting isometries.
We now discuss this $O(2)$ subgroup in more detail.

The ele\-ven-di\-men\-sion\-al theory is obviously invariant under the
group $O(2) = SO(2) \times \Z_{2}$ of rotations and reflections in the
$xy$ plane\footnote{Observe that $\cal{C}$ transforms as a pseudotensor,
and, therefore, changes sign under reflections in the $xy$ plane.},
inducing an $O(2)$ invariance of the ni\-ne-di\-men\-sion\-al theory.
The infinitesimal form of the $SO(2)$ transformations of the scalars and
vector fields is

\begin{equation}
\begin{array}{rclrcl}
\delta k & = & \frac{1}{2} \theta k\ell\, ,
\hspace{1cm}&
\delta A^{(1)}
&
=
&
-\theta A^{(2)}\, ,
\\
& & & & & \\
\delta e^{\phi}
&
=
&
-\frac{7}{4}\theta \ell e^{\phi}\, ,
&
\delta A^{(2)} & = & \theta A^{(1)}\, ,
\\
& & & & & \\
\delta \ell
&
=
&
\theta \left( 1 +\ell^{2} -2k e^{-2\phi}\right)\, ,
\hspace{1.5cm}
&
\delta B
&
=
&
0\, ,
\\
\end{array}
\label{eq:o21}
\end{equation}

\noindent and those of the remaining fields are

\begin{equation}
\begin{array}{rclrcl}
\delta B^{(1)}
&
=
&
-\theta B^{(2)}\, , \hspace{2cm}
&
\delta B^{(2)}
&
=
&
\theta B^{(1)}\, ,
\\
& & & & & \\
\delta g_{\mu\nu}
&
=
&
-\theta \ell g_{\mu\nu} \, ,
&
\delta C
&
=
&
0\, ,
\end{array}
\label{eq:o22}
\end{equation}

\noindent where $\theta$ is an infinitesimal constant parameter.  On the
other hand, the discrete $\Z_{2}$ transformations, corresponding to the
reflection $y\rightarrow -y$, is given by:

\begin{equation}
\begin{array}{rclrcl}
\ell^{\prime}
&
=
&
-\ell\, , \hspace{1.5cm}
&
A^{(1)\prime}
&
=
&
- A^{(1)}\, ,
\\
& & & & & \\
B^{(2)\prime}
&
=
&
- B^{(2)}\, ,
&
C^{\prime}
&
=
& -C\, ,
\\
\end{array}
\label{eq:z2}
\end{equation}

\noindent and the remaining fields are invariant.  A particularly
interesting $O(2)$--rotated version of this $\Z_2$--transformation is
given by an interchange of the coordinates $x$ and $y$\footnote{This
transformation corresponds to a finite $O(2)$ rotation with parameter
$\theta =-\frac{\pi}{2}$ followed by the reflection $y \rightarrow
-y$.}, under which the ni\-ne-di\-men\-sion\-al scalars and vectors
transform as follows

\begin{equation}
\begin{array}{rclrcl}
k^{\prime} & = &
k \left( \ell^{2} +ke^{-2\phi}\right)^{-\frac{1}{4}}\, ,
\hspace{1.2cm}&
A^{(1)\prime} & = & A^{(2)}\, ,
\\
& & & & &
\\
\ell^{\prime} & = &
\ell \left( \ell^{2} +ke^{-2\phi}\right)^{-1}\, ,&
A^{(2)\prime} & = & A^{(1)}\, ,
\\
& & & & &
\\
e^{\phi^{\prime}} & = &
e^{\phi} \left( \ell^{2} +ke^{-2\phi}\right)^{\frac{7}{8}}\, ,&
B^{\prime} & = & B\, ,
\\
\end{array}
\label{eq:xy1}
\end{equation}

\noindent and the remaining fields

\begin{equation}
\begin{array}{rclrcl}
B^{(1)\prime} & = & -B^{(2)}\, , &
B^{(2)\prime} & = & -B^{(1)}\, ,
\\
& & & & &
\\
g_{\mu\nu}^{\prime} & = & \left( \ell^{2}
+ke^{-2\phi} \right)^{\frac{1}{2}} g_{\mu\nu}\, ,
\hspace{1cm}&
C^{\prime} & = & -C\, .
\end{array}
\label{eq:xy2}
\end{equation}

We now consider the ten-di\-men\-sion\-al reformulation of these
symmetries.  The ni\-ne-di\-men\-sion\-al $O(2)$ invariance
Eqs.~(\ref{eq:o21},\ref{eq:o22},\ref{eq:z2}) corresponds to non-trivial
dualities of both ten-di\-men\-sion\-al type--II supergravity theories.
As an example of this kind of duality we write down the
ten-di\-men\-sion\-al type--II transformations corresponding to the
finite $\Z_2$--transformations given in
Eqs.~(\ref{eq:xy1},\ref{eq:xy2}).  We will provisionally call this
$\Z_2$ transformation a type--II ``$xy$--duality''.  The explicit form
of the type--IIA $xy$--duality rules is given by:

\begin{equation}
\begin{array}{rclrcl}
\hat{\phi}^{\prime} & = & \hat{\phi} + \frac{3}{4}
\log\hat{G}_{\underline{x}\underline{x}}\, ,
&
\hat{A}^{(1)\prime}_{\underline{x}}
&
=
&
\hat{A}^{(1)}_{\underline{x}}
\hat{G}_{\underline{x}\underline{x}}^{-1}\, ,
\\
& & & & & \\
\hat{B}^{(1)\prime}_{\mu\nu}
&
=
&
-\frac{3}{2} \hat{C}_{\mu\nu\underline{x}}\, ,
&
\hat{A}^{(1)\prime}_{\mu}
&
=
&
\left( \hat{A}^{(1)}_{\underline{x}} \hat{A}^{(1)}_{\mu}
-e^{-2\hat{\phi}}\hat{g}_{\mu\underline{x}} \right)
\hat{G}_{\underline{x}\underline{x}}^{-1}\, ,
\\
& & & & & \\
\hat{B}^{(1)\prime}_{\underline{x}\mu}
&
=
&
\hat{B}^{(1)}_{\underline{x}\mu}\, ,
&
\hat{g}_{\underline{x}\underline{x}}^{\prime}
&
=
&
\hat{g}_{\underline{x}\underline{x}}
\hat{G}_{\underline{x}\underline{x}}^{-\frac{1}{2}}\, ,
\\
& & & & & \\
\hat{g}_{\mu\underline{x}}^{\prime}
&
=
&
2\hat{A}^{(1)}_{[\mu} \hat{g}_{\underline{x} ]\underline{x}}
\hat{G}_{\underline{x}\underline{x}}^{-\frac{1}{2}}\, ,
\hspace{1cm}
&
\hat{g}_{\mu\nu}^{\prime}
&
=
&
-e^{2\hat{\phi}} \hat{G}_{\underline{x}\underline{x}}^{-\frac{1}{2}}
\left( \hat{G}_{\underline{x}\underline{x}} \hat{G}_{\mu\nu}
-\hat{G}_{\mu\underline{x}} \hat{G}_{\nu\underline{x}}
\right)
\\
& & & & & \\
\hat{C}^{\prime}_{\mu\nu\rho}
&
=
&
-\hat{C}_{\mu\nu\rho}\, ,
&
\hat{C}^{\prime}_{\mu\nu\underline{x}}
&
=
&
-\frac{2}{3} \hat{B}^{(1)}_{\mu\nu}\, ,
\\
\label{eq:xyIIA}
\end{array}
\end{equation}

\noindent where

\begin{equation}
\hat{G}_{\hat{\mu}\hat{\nu}} = \hat{A}_{\hat{\mu}}^{(1)}
\hat{A}_{\hat{\nu}}^{(1)}
-e^{-2\hat{\phi}}\hat{g}_{\hat{\mu}\hat{\nu}}\, .
\end{equation}

Similarly, the type--IIB $xy$--duality transformations are given by:

\begin{equation}
\begin{array}{rclrcl}
\hat{D}_{\mu\nu\rho\underline{x}}^{\prime}
&
=
&
-\hat{D}_{\mu\nu\rho\underline{x}}\, ,
&
\hat{\jmath}_{\mu\nu}^{\prime}
&
=
&
|\hat{\lambda}|\hat{\jmath}_{\mu\nu}\, ,
\\
& & & & & \\
\hat{\jmath}_{\underline{x}\mu}^{\prime}
&
=
&
|\hat{\lambda}|\hat{\jmath}_{\underline{x}\mu}\, ,
&
\hat{\jmath}_{\underline{x}\underline{x}}^{\prime}
&
=
&
|\hat{\lambda}|\hat{\jmath}_{\underline{x}\underline{x}}\, ,
\\
& & & & & \\
\hat{\cal B}^{(1)\prime}_{\mu\nu}
&
=
&
-\hat{\cal B}^{(2)}_{\mu\nu}\, ,
&
\hat{\cal B}^{(2)\prime}_{\mu\nu}
&
=
&
-\hat{\cal B}^{(1)}_{\mu\nu}\, ,
\\
& & & & & \\
\hat{\cal B}^{(1)\prime}_{\underline{x}\mu}
&
=
&
-\hat{\cal B}^{(2)}_{\underline{x}\nu}\, ,
&
\hat{\cal B}^{(2)\prime}_{\underline{x}\nu}
&
=
&
-\hat{\cal B}^{(1)}_{\underline{x}\nu}\, ,
\\
& & & & & \\
\hat{\ell}^{\prime}
&
=
&
|\hat{\lambda}|^{-2}\hat{\ell}\, ,\hspace{2cm}
&
\hat{\varphi}^{\prime}
&
=
&
\hat{\varphi} +2\log{|\hat{\lambda}|}\, ,
\end{array}
\label{eq:xyIIB}
\end{equation}

\noindent (recall that $\hat\lambda = \hat\ell + ie^{-\hat\varphi}$).

Observe that the $xy$--dualities interchange (and mix) NS-NS fields with
RR ones, and can be used to generate solutions with non-trivial RR
fields from solutions of the NS-NS sector (which are also solutions of
the heterotic string with no background gauge fields).  In the type--IIB
theory the $xy$--duality transformations Eqs.~(\ref{eq:xyIIB}) is the
$S$--duality transformation under which

\begin{equation}
\hat{\lambda}^{\prime}=-1/\hat{\lambda}\, ,
\end{equation}

\noindent combined with other discrete symmetries of the theory.

In the case of the type--IIA theory, the $xy$--duality has its origin in
the $O(2)$ symmetry of the ele\-ven-di\-men\-sion\-al theory restricted
to backgrounds with two commuting isometries.

The type--IIA theory, when restricted to backgrounds with one isometry,
has an $SL(2,\R)$ $S$--duality invariance which includes the
$xy$--duality Eqs.~(\ref{eq:xyIIA}).  Note that if we set
$\hat{A}^{(1)}=0,\,\, \hat{g}_{\underline{x}\underline{x}}=-1$ in
Eqs.~(\ref{eq:xyIIA}) (for simplicity) then the type--IIA $xy$--duality
transformation relates the strong- and weak-coupling regimes of the
underlying type--IIA superstring theory:

\begin{equation}
\hat{\phi}^{\prime}=-{\textstyle\frac{1}{2}}\hat{\phi}\, ,
\end{equation}

\noindent Note also that Eqs.~(\ref{eq:xyIIA}) and Eqs.~(\ref{eq:xyIIB})
are related by a type--II $T$--duality transformation as will be
discussed below.

We now consider the construction of the type--II $T$--duality rules.  It
turns out that the derivation of these rules is rather subtle since the
type--II $T$--transformations do not correspond to a non-compact
symmetry of the ni\-ne-di\-men\-sion\-al theory.  As mentioned in the
introduction, this is related to the fact that the type--II $T$ duality
maps one theory (the type--IIA superstring) onto another theory (the
type--IIB superstring).  Consider first the NS-NS truncation of the
ni\-ne-di\-men\-sion\-al theory, with the type--I action:

\begin{eqnarray}
S  & = &
{\textstyle\frac{1}{2}}\int d^{9}x\  \sqrt{g}\   e^{-2\phi}
\left[ -R +4(\partial\phi)^{2} -{\textstyle\frac{3}{4}}
\left( H^{(1)} \right)^{2} \right.
\nonumber \\
& &
\nonumber \\
& &
\left. -(\partial\log k)^{2} +{\textstyle\frac{1}{4}}k^{2}\left( F^{(2)}
\right)^{2} +{\textstyle\frac{1}{4}} k^{-2} F^{2}(B)\right]\, .
\label{eq:9dactionns}
\end{eqnarray}

This has an $O(1,1) = SO(1,1) \times \Z_2$ duality symmetry.  The
ni\-ne-di\-men\-sion\-al $\Z_2$--transformation is given by:

\begin{equation}
\tilde A_\mu^{(2)} = B_\mu\, ,\hskip 1truecm
\tilde B_\mu = A_\mu^{(2)}\, ,\hskip 1truecm
\tilde k = k^{-1}\, .
\label{eq:Ztwo}
\end{equation}

\noindent This is the standard $T$--duality transformation
\cite{kn:Bu1}.  (Note that $k$ is the modulus field for the
compactifying circle, so that its expectation value corresponds to the
radius $R$.)  The continuous $SO(1,1)$ symmetry scales $k$ and acts by

\begin{equation}
\tilde k = \Lambda k \,
 ,\hskip 1truecm
\tilde B_\mu =\Lambda B_\mu \, ,\hskip 1truecm
\tilde A_\mu^{(2)} = \Lambda^{-1} A_\mu^{(2)} \, .
\end{equation}

\noindent This corresponds to a particular general coordinate
transformation in $d=10$\footnote{It is not always the case that the
a continuous transformation of a $T$--duality group is a particular
gauge transformation in a higher dimensional theory.  The simplest
counter-example is provided by considering the coupling of the type--I
string to one Abelian vector multiplet.  The $T$--duality symmetry in
nine dimensions is extended from $O(1,1)$ to $O(2,1)$. $SO(2,1)$ has
several discrete transformations that take us from the sheet of $O(2,1)$
which is connected to the identity to other sheets.  Each of them
generates a $\Z_{2}$ subgroup.  One of them is Buscher's $T$--duality.
Each sheet of $O(2,1)$, and, in particular, the one connected to the
identity, is three-dimensional: one transformation is a special
g.c.t.~transformation in $d=10$, another corresponds to a special $U(1)$
gauge transformation in $d=10$ but the third one yields a non-trivial
solution-generating transformation in $d=10$ \cite{kn:Sen1}.  The effect
of this transformation is to convert uncharged solutions into charged
ones.  For more details about this case, see \cite{kn:Be6}.}.

The $SO(1,1)$ transformations extend to a symmetry of the full $d=9$
type--II action (\ref{eq:9daction}) under which each field $A$ scales
with some weight $w$: $ A \to \Lambda ^w A$.  The weights of the fields
are given in Table~\ref{tab-weightsb}.

\begin{table}
\begin{center}
\begin{tabular}{||l|c|l|c||}
\hline
field&weight $w$&field&weight $w$\\
\hline
& & & \\
$k$&1&$B$&1\\
& & & \\
${  A}^{(1)}$&-1/2&${  A}^{(2)}$&-1 \\
& & & \\
${B}^{(1)}$&0&${B}^{(2)}$&1/2\\
& & & \\
$\ell$&1/2&$  C$&-1/2\\
& & & \\
\hline
\end{tabular}
\end{center}

\caption{This table gives the weights $w$ of the fields of $d=9$
type--II supergravity under the global $SO(1,1)$ symmetry.}
\label{tab-weightsb} \end{table}

However, the $\Z_2$ transformations (\ref{eq:Ztwo}) do not extend to any
symmetry of the $d=9$ action.  Thus the $T$--duality transformations
relating type--IIA backgrounds to type--IIB ones cannot be found from
symmetries of the $d=9$ theory.  Instead, we find the type--II
$T$--duality rules as follows.  As we have seen in the previous
sections, the compactification of both the ten-di\-men\-sion\-al
type--IIA and type--IIB theories lead to the same
ni\-ne-di\-men\-sion\-al supergravity theory.  Therefore, the same
ni\-ne-di\-men\-sion\-al field configuration can be embedded in a
ten-di\-men\-sion\-al theory (or \lq decompactified') in two different
ways\footnote{ One way of embedding is given in (\ref{eq:basis2}) while
the other way is given in (\ref{eq:basis3}).} yielding two different
ten-di\-men\-sion\-al field configurations of two different theories.

Using the two inequivalent embeddings given in Eqs.~(\ref{eq:basis2})
and Eqs.~(\ref{eq:basis3}) one finds that the transformation rules for
the type--II duality symmetry that maps the type--IIB superstring onto
the type--IIA superstring is given by:

\begin{equation}
\begin{array}{rclrcl}
\hat{C}_{\underline{x}\mu\nu} & = &
{\textstyle\frac{2}{3}}\left[ \hat{\cal B}_{\mu\nu}^{(2)}
+2 \hat{\cal B}^{(2)}_{\underline{x}[\mu}\hat{\jmath}_{\nu]\underline{x}}
/\hat{\jmath}_{\underline{x}\underline{x}}
\right]\, ,
& & & \\
& & & & & \\
\hat{C}_{\mu\nu\rho} & = &
{\textstyle\frac{8}{3}} \hat{D}_{\underline{x}\mu\nu\rho}
+\epsilon^{ij}\hat{\cal B}^{(i)}_{\underline{x}[\mu}
\hat{\cal B}^{(j)}_{\nu\rho]}
+\epsilon^{ij}\hat{\cal B}^{i}_{\underline{x}[\mu}
\hat{\cal B}^{j}_{|\underline{x} |\nu} \hat{\jmath}_{\rho]\underline{x}}
/\hat{\jmath}_{\underline{x}\underline{x}}\, ,
\hspace{-3cm}
& & & \\
& & & & & \\
\hat{g}_{\mu\nu}
&
=
&
\hat{\jmath}_{\mu\nu}
-\left( \hat{\jmath}_{\underline{x}\mu}\hat{\jmath}_{\underline{x}\nu}
-\hat{\cal B}^{(1)}_{\underline{x}\mu}
\hat{\cal B}^{(1)}_{\underline{x}\nu} \right)
/\hat{\jmath}_{\underline{x}\underline{x}}\, ,
& & & \\
& & & & & \\
\tilde{B}^{(1)}_{\mu\nu}
&
=
&
\hat{\cal B}^{(1)}_{\mu\nu}
+2\hat{\cal B}^{(1)}_{\underline{x}[\mu}
\hat{\jmath}_{\nu]\underline{x}}
/\hat{\jmath}_{\underline{x}\underline{x}}\, ,
& & & \\
\end{array}
\end{equation}

\begin{equation}
\begin{array}{rclrcl}
\hat{g}_{\underline{x}\mu}
&
=
&
\hat{\cal B}_{\underline{x}\mu}^{(1)}
/\hat{\jmath}_{\underline{x}\underline{x}}\, ,
&
\hat{B}_{\underline{x}\mu}^{(1)}
&
=
&
\hat{\jmath}_{\underline{x}\mu}
/\hat{\jmath}_{\underline{x}\underline{x}}
\\
& & & & & \\
\hat{A}^{(1)}_{\mu}
&
=
&
-\hat{\cal B}_{\underline{x}\mu}^{(2)} +\hat{\ell}
\hat{\cal B}_{\underline{x}\mu}^{(1)}\, ,
&
\hat{g}_{\underline{x}\underline{x}}
&
=
&
1/\hat{\jmath}_{\underline{x}\underline{x}}\, ,
\\
& & & & & \\
\hat{\phi}
&
=
&
\hat{\varphi} -\frac{1}{2}
\log{(-\hat{\jmath}_{\underline{x}\underline{x}})}\, ,
&
\hat{A}_{\underline{x}}^{(1)}
&
=
&
\hat{\ell}\, .
\\
\end{array}
\label{eq:TBA}
\end{equation}

\noindent Similarly, the type--II duality map from the type--IIA onto
the type--IIB superstring is given by:

\begin{eqnarray}
\hat{D}_{\underline{x}\mu\nu\rho}
&
=
&
{\textstyle\frac{3}{8}}\left[
\hat{C}_{\mu\nu\rho} -\hat{A}^{(1)}_{[\mu}\hat{B}^{(1)}_{\nu\rho]}
+\hat{g}_{\underline{x}[\mu}\hat{B}^{(1)}_{\nu\rho]}
\hat{A}^{(1)}_{\underline{x}}/\hat{g}_{\underline{x}\underline{x}}
-{\textstyle\frac{3}{2}}\hat{g}_{\underline{x}[\mu}
\hat{C}_{\nu\rho]\underline{x}}
/\hat{g}_{\underline{x}\underline{x}}\right]\, ,
\nonumber \\
& &
\nonumber \\
\hat{\jmath}_{\mu\nu}
&
=
&
\hat{g}_{\mu\nu} -\left(\hat{g}_{\underline{x}\mu}
\hat{g}_{\underline{x}\nu}
-\hat{B}^{(1)}_{\underline{x}\mu}
\hat{B}^{(1)}_{\underline{x}\nu}\right)
/\hat{g}_{\underline{x}\underline{x}}\, ,
\nonumber \\
& &
\nonumber \\
\hat{\cal B}^{(2)}_{\mu\nu}
&
=
&
{\textstyle\frac{3}{2}}\hat{C}_{\mu\nu\underline{x}}
-2 \hat{A}^{(1)}_{[\mu}\hat{B}^{(1)}_{\nu]\underline{x}}
+2\hat{g}_{\underline{x}[\mu}\hat{B}^{(1)}_{\nu]\underline{x}}
\hat{A}^{(1)}_{\underline{x}}/\hat{g}_{\underline{x}\underline{x}}\, ,
\nonumber
\end{eqnarray}

\begin{equation}
\begin{array}{rclrcl}
\hat{\cal B}^{(1)}_{\mu\nu}
&
=
&
\hat{B}^{(1)}_{\mu\nu} +2\hat{g}_{\underline{x}[\mu}
\hat{B}^{(1)}_{\nu]\underline{x}}
/\hat{g}_{\underline{x}\underline{x}}\, ,
&
\hat{\jmath}_{\underline{x}\mu}
&
=
&
\hat{B}^{(1)}_{\underline{x}\mu}
/\hat{g}_{\underline{x}\underline{x}}\, ,
\\
& & & & & \\
\hat{\cal B}^{(1)}_{\underline{x}\mu}
&
=
&
\hat{g}_{\underline{x}\mu}/\hat{g}_{\underline{x}\underline{x}}\, ,
&
\hat{\cal B}^{(2)}_{\underline{x}\mu}
&
=
&
-\hat{A}^{(1)}_{\mu} +\hat{A}^{(1)}_{\underline{x}}
\hat{g}_{\underline{x} \mu}/\hat{g}_{\underline{x}\underline{x}}\, ,
\\
& & & & & \\
\hat{\jmath}_{\underline{x}\underline{x}}
&
=
&
1/\hat{g}_{\underline{x}\underline{x}}\, ,
&
\hat{\ell}
&
=
&
\hat{A}_{\underline{x}}^{(1)}\, ,
\\
& & & & & \\
\hat{\varphi}
&
=
&
\hat{\phi}-\frac{1}{2}\log{(-\hat{g}_{\underline{x}\underline{x}})}\, .
& & & \\
\end{array}
\label{eq:TAB}
\end{equation}

The dual of the type--IIA metric $\hat{g}_{\underline{x}\underline{x}}$
is given by the inverse of the type--IIB metric
$\hat{\jmath}_{\underline{x}\underline{x}}$ and vice versa.  For a torus
compactification this means that the usual $R\rightarrow
\alpha^\prime/R$ duality is replaced by the map $R_{IIA} \rightarrow
\alpha^\prime/R_{IIB}$ where $R_{IIA}$ is the torus radius
characterizing the type--IIA decompactification and $R_{IIB}$ is the
torus radius characterizing the type--IIB decompactification, as in
Refs.~\cite{kn:Da1,kn:Di1}.

We observe that the type--II $T$--duality rules are a true
generalization of Buscher's duality rules \cite{kn:Bu1} in the sense
that if we set the type--IIA and type--IIB Ramond--Ramond fields to zero
and identify the remaining NS-NS type--IIA and type--IIB fields with
the type--I fields, the above rules reduce to (\ref{eq:Ztwo}).
Furthermore, note that the type--II duality rule is a a non-trivial
solution-generating transformation in the following sense: given a
solution of the type--IIA string equations of motion with one isometry,
it generates a solution of the type--IIB equations of motion and vice
versa.

This type--II $T$--duality maps the symmetries of each individual
ten-di\-men\-sion\-al type--II theory into the other. This is
specially useful when one symmetry is manifest in one theory but not
in the other. This is the case of $SL(2,\R)$, which is manifest in the
type--IIB theory (with or without isometries) but it is not manifest
by any means in the type--IIA theory (with one isometry).The reader can
check that the type--IIB $S$--duality rules Eqs.~(\ref{eq:xyIIB}) are
mapped into the type--IIA $S$--duality rules Eqs.~(\ref{eq:xyIIA}) by
the type--II $T$--duality rules Eqs.~(\ref{eq:TBA}).

The analysis we have given for the bosonic sector can be
straightforwardly extended to the full supersymmetric theory with
fermions, since the non-compact symmetries of the bosonic sector are
known to extend to symmetries of the full supergravity theory.  Of
particular interest are supersymmetric solutions which admit Killing
spinors, and we now address the question of whether the image of a
supersymmetric solution under duality is again supersymmetric.  For
example, the $xy$--duality transformations are simple coordinate
transformations in eleven dimensions and, therefore, they preserve
ele\-ven-di\-men\-sion\-al unbroken supersymmetries.  If the
eleven-di\-men\-sio\-nal Killing spinors corresponding to a given
solution are {\sl independent} of the coordinates $x$ and $y$, they will
be invariant under this duality transformation.  Under these conditions,
upon compactification of the coordinates $x$ or $y$ or a combination of
both, we will get ten-di\-men\-sion\-al unbroken supersymmetries.  The
Killing spinors will depend on which coordinate we have compactified and
the different choices will be related by $xy$--duality transformations
in ten dimensions.  On the other hand, if the ele\-ven-di\-men\-sion\-al
Killing spinors depend on $x$ or $y$ we expect that supersymmetry will
be broken by $xy$--duality, as in the case studied in
Ref.~\cite{kn:Be5}.  We have seen that the type--II $T$--duality rules
do not correspond to any symmetry at all in nine dimensions.  Therefore,
all ni\-ne-di\-men\-sion\-al properties will be preserved, in particular
unbroken supersymmetries.  Again everything depends on the preservation
of the Killing spinors in the compactification procedure.
Ten--dimensional Killing spinors with explicit dependence on the
direction with respect to which we are going to dualize will lead to
broken supersymmetry while duality will commute with the spacetime
supersymmetry if the Killing spinors are independent of the duality
direction.

The type--IIA $S$--duality rules are based on the existence of two
isometries corresponding to the directions $x$ and $y$.  It is
interesting to note that transformations based on the existence of two
isometries in the higher-dimensional theory have been considered before,
albeit in a slightly different context, in the construction of the
Kaluza-Klein or Gross-Perry-Sorkin (GPS) magnetic monopole
\cite{kn:GPS}\footnote{We note that recently a six-brane solution of
ele\-ven-di\-men\-sion\-al supergravity has been constructed which is an
exact analogue in eleven dimensions of the GPS monopole in five
dimensions \cite{kn:To1}.}.  In essence, in the GPS case one considers a
five-dimensional configuration with two isometries and ``compactifies"
alternatively the two corresponding directions getting two
four-dimensional configurations each of them with a different isometry
(the Euclidean Taub-NUT solution and the GPS magnetic monopole).

In our language we could say that these two configurations are {\it
dual}.  There are only a few non-essential differences between the GPS
case and our case:

\begin{enumerate}

\item The original higher dimensional theory.

\item The fact that in the GPS case one of the isometry directions is
time-like and the other one is space-like while in our case both
isometry directions are space-like.  The compactification of a time-like
direction leads to a four-dimensional Euclidean Kaluza-Klein theory with
a vector field and a scalar.  In order to avoid the occurrence of the
vector field one has to impose more restrictive conditions on the
higher-dimensional configurations: they must be not just
time-independent (stationary) but {\it static}\footnote{The time-like
Killing vector is then ``hypersurface-orthogonal" which in practice
means that all the elements $g^{(5)}_{\underline{0}\underline{i}}$ of
the five-dimensional metric can be made to vanish in an appropriate
coordinate system}.  The presence of the unwanted scalar can be avoided
by choosing five-dimensional configurations as those considered in
Refs.~\cite{kn:GPS} with $g^{(5)}_{\underline{0}\underline{0}} =1$.

\end{enumerate}


\section{Duality In Eleven Dimensions}
\label{sec-elevendual}

The ele\-ven-di\-men\-sion\-al supergravity theory has no duality
symmetries of its equations of motion for general backgrounds.  For
backgrounds with one isometry, there should be an $SO(1,1)$ symmetry of
the equations of motion corresponding to the $S$--duality of the
type--IIA theory; this is essentially a particular
ele\-ven-di\-men\-sion\-al diffeomorphism.  For backgrounds with two
isometries, there should be an $SO(1,1)\times SL(2,\R)$ symmetry of the
equations of motion corresponding to the duality symmetries of the $d=9$
theory.  We have already identified an $O(2)$ subgroup of $SL(2,\R)$ as
rotations and reflections in the $xy$ plane.  It is clear that there
cannot be an analogue of the $\Z_2$ $T$--duality symmetry here as the
type--IIB supergravity theory cannot be obtained from any eleven
dimensional theory.  However, if we restrict ourselves to the {\it
subset} of solutions of $N=1,d=11$ supergravity which have two commuting
isometries in the directions parametrized by the coordinates
$y=x^{\underline{10}}$ and $x=x^{\underline{9}}$ and which, in addition,
satisfy

\begin{equation}
\hat{\hat{C}}_{\hat{\mu}\hat{\nu}\hat{\rho}}
=\hat{\hat{g}}_{\hat{\mu}\underline{y}}=0\, .
\end{equation}

\noindent then the configuration of $N=1,d=11$ gives a solution of
$N=1,d=10$ supergravity with one isometry upon dimensional reduction,
and this has a $\Z_2$ Buscher duality symmetry.  The algebraic
constraints are then the truncation from type--IIA supergravity to
$N=1,d=10$ supergravity Eq.~(\ref{eq:truncationIIA}) written in eleven
dimensions and the $T$--duality rules can be rewritten in
ele\-ven-di\-men\-sion\-al form:

\begin{equation}
\begin{array}{rclrcl}
\hat{\hat{C}}_{\mu\nu\underline{y}}^{\prime}
&
=
&
\hat{\hat{C}}_{\mu\nu\underline{y}}
+2\hat{\hat{g}}_{\underline{x}[\mu}
\hat{\hat{C}}_{\nu]\underline{x}\underline{y}}
/\hat{\hat{g}}_{\underline{x}\underline{x}}\,
&
\hat{\hat{C}}_{\mu\underline{x}\underline{y}}^{\prime}
&
=
&
-\frac{2}{3}\hat{\hat{g}}_{\underline{x}\mu}
/\hat{\hat{g}}_{\underline{x}\underline{x}}\, ,
\\
& & & & & \\
\hat{\hat{g}}_{\mu\nu}^{\prime}
&
=
&
\left(-\hat{\hat{g}}_{\underline{y}\underline{y}} \right)^{\frac{1}{6}}
\left(-\hat{\hat{g}}_{\underline{x}\underline{x}} \right)^{\frac{1}{3}}
\left\{ \hat{\hat{g}}_{\mu\nu} -\hat{\hat{g}}_{\underline{x}\mu}
\hat{\hat{g}}_{\underline{x}\nu}
/\hat{\hat{g}}_{\underline{x}\underline{x}}
-\frac{9}{4}\hat{\hat{C}}_{\mu\underline{x}\underline{y}}
\hat{\hat{C}}_{\nu\underline{x}\underline{y}}
/\left( \hat{\hat{g}}_{\underline{x}\underline{x}}
\hat{\hat{g}}_{\underline{y}\underline{y}}\right)\right\}\, ,
\hspace{-10cm}
& & & \\
& & & & & \\
\hat{\hat{g}}_{\underline{x}\mu}^{\prime}
&
=
&
\frac{3}{2}
\left(-\hat{\hat{g}}_{\underline{y}\underline{y}}\right)^{-\frac{5}{6}}
\left(-\hat{\hat{g}}_{\underline{x}\underline{x}}\right)^{-\frac{2}{3}}
\hat{\hat{C}}_{\mu\underline{x}\underline{y}}\, ,
&
\hat{\hat{g}}_{\underline{y}\underline{y}}^{\prime}
&
=
&
-\left(-\hat{\hat{g}}_{\underline{y}\underline{y}}\right)^{\frac{2}{3}}
\left(-\hat{\hat{g}}_{\underline{x}\underline{x}}\right)^{-\frac{2}{3}}\, ,
\\
& & & & & \\
\hat{\hat{g}}_{\underline{x}\underline{x}}^{\prime}
&
=
&
-\left(-\hat{\hat{g}}_{\underline{y}\underline{y}}\right)^{-\frac{5}{6}}
\left(-\hat{\hat{g}}_{\underline{x}\underline{x}}\right)^{-\frac{2}{3}}\, .
& & & \\
\end{array}
\label{eq:T11}
\end{equation}

The condition $\hat{\hat{g}}_{\hat{\mu}\underline{y}}=0$ means that the
Killing vector $ \partial / \partial y$ is hypersurface--orthogonal,
i.e. orthogonal to the hypersurfaces of constant value of $y$.  The
ele\-ven-di\-men\-sion\-al manifold ${\cal M}^{11}$ is the product of a
ten-di\-men\-sion\-al manifold times a circle ${\cal M}^{11} ={\cal
M}^{10}\times S^{1}$.

It is interesting to see what the membrane analogue is of the usual $R
\rightarrow 1/ R$ duality\footnote{For simplicity, we assume from now
on that all radii and fields have been redefined to be dimensionless, as
in \cite{kn:Gi1}.}.  For this purpose we consider a membrane moving in
the space ${\cal M}^9 \times T^2$ (ni\-ne-di\-men\-sion\-al Minkowski
space times a two-torus) and assume that the radius of the two-torus
in the $ x^{\underline 9}=x$--direction is $R_1$, i.e.~${\hat {\hat
g}}_{\underline {xx}} = -(R_1)^2$ and similarly that the radius in the
$x^{\underline {10}}=y$--direction is $R_2$, i.e.~${\hat {\hat
g}}_{\underline {yy}} = -(R_2)^2$.  We find that for this case the
duality rules are given by:

\begin{eqnarray}
R_1^\prime &=& 1/\left (R_1^{2/3}R_2^{5/6}\right )\, ,\hskip 1.5truecm
R_2^\prime = \left ( R_2/R_1 \right )^{2/3}\, ,\nonumber\\
& & \nonumber \\
{\hat{\hat \eta}}_{\mu\nu}^\prime & = & R_1^{2/3} R_2^{1/3}\
{\hat{\hat\eta}}_{\mu\nu}\, .
\end{eqnarray}

It is well known that in case of the string duality the one-torus with
the self-dual radius $R=1$ is special in the sense that
symmetry-enhancement occurs.  We find that in the case of the membrane
there is a whole one-parameter family of two-tori which are self-dual.
They are characterized by the following radii:

\begin{equation}
R_1 = R\, ,\hskip 2truecm R_2 = {1\over R^2}\, .
\end{equation}

\noindent It would be interesting to see in which sense this family of
two tori plays a special role in membrane dynamics.


\section{Examples}
\label{sec-examples}

As an illustration of our results we shall now apply the duality
transformations constructed in previous sections to generate new
solutions of the type--IIA, type--IIB and ele\-ven-di\-men\-sion\-al
supergravity theories.  Our starting point will be the ``Supersymmetric
String Waves" ($SSW$) of Ref.~\cite{kn:Be3} which are solutions of the
heterotic string and also of the type--II equations of motion.  Under
type--I $T$--duality they are dual to the ``Generalized Fundamental
Strings" ($GFS$) solutions of Refs.~\cite{kn:Be4,kn:Dab1}.

Both the $SSW$ and $GFS$ solutions can be embedded into the type--I,
type--IIA and type--IIB theories.  We will denote the embedded solutions
by $SSW,SSW(A)$ and $SSW(B)$ respectively and similarly for the $GFS$.
We start with $SSW(A)$ and $GFS(A)$ and we first perform a discrete
$xy$--duality transformation using Eqs.~(\ref{eq:xyIIA}).  The
$xy$--duality generates new solutions of the type--IIA equations of
motion which we denote by $SSW(A^\prime)$ and $GFS(A^\prime)$,
respectively.  Next, we perform a type--II $T$--duality transformation
to the type--IIB theory according to Eqs.~(\ref{eq:TAB}).  This leads to
new solutions of the type--IIB theory where $SSW$ solutions are
converted into $GFS$ solutions and vice--versa.  We denote these new
solutions by $GFS(B^{\prime})$ and $SSW(B^{\prime})$, respectively.
Finally, we perform a further $xy$--duality transformation using
Eqs.~(\ref{eq:xyIIB}) getting $GFS(B)$ and $SSW(B)$.  The reader may
check that the $GFS(B)$ and $SSW(B)$ solutions are related by the
type--II $T$--duality Eqs.~(\ref{eq:TBA}) to the original $SSW(A)$ and
$GFS(A)$ solutions we started from, as they should.  Below we give the
explicit form of the new solutions obtained in this manner.


\subsection{Duality rotation of SSW}

We first consider the SSW case. The fields of the SSW(A)
solution are given by:

\begin{equation}
SSW(A)
\left\{
\begin{array}{rcl}
ds^{2}
&
=
&
2\left(dv +{\cal A}_{u}du +2{\cal A}_{\underline{i}}dx^{\underline{i}}
\right)du -dx^{\underline{i}}dx^{\underline{i}}\, ,
\\
& & \\
\hat{B}^{(1)}
&
=
&
2 {\cal A}_{\underline{i}}dx^{\underline{i}}\wedge du\, ,
\\
& & \\
\hat{\phi}
&
=
&
0\, .
\end{array}
\right.
\end{equation}
The indices $i,j$ run from $1$ to $8$ and
$u=\frac{1}{\sqrt{2}}(t+x),\,\, v=\frac{1}{\sqrt{2}}(t-x)$.
Here ${\cal A}_u$ and ${\cal A}_{\underline i}$ are arbitrary
functions, independent of $u$ and $v$, that satisfy the equations

\begin{equation}
\triangle {\cal A}_u = 0\, ,\hskip 1.5truecm
\triangle \partial^{[\underline i} {\cal A}^{{\underline {j}}]}=0\, ,
\end{equation}

\noindent where the Laplacian is taken over the eight transverse
directions only.

Performing the $xy$--duality transformations Eqs.~(\ref{eq:xyIIA})
we get the new $SSW(A^\prime)$ solution:

\begin{equation}
SSW(A^{\prime})
\left\{
\begin{array}{rcl}
ds^{2}
&
=
&
e^{-\frac{2}{3}\hat{\phi}}
\left( dt +{\textstyle\frac{1}{\sqrt{2}}}
{\cal A}_{\underline{i}}dx^{\underline{i}}\right)^{2}
-e^{-\frac{2}{3}\hat{\phi}}
\left( dx^{2} +dx^{\underline{i}}dx^{\underline{i}}\right)
\\
& & \\
\hat{B}^{(1)}
&
=
&
-{\textstyle\frac{1}{\sqrt{2}}}
{\cal A}_{\underline{i}} dx^{\underline{i}} \wedge dx\, ,
\\
& & \\
\hat{C}
&
=
&
{\textstyle\frac{\sqrt{2}}{3}}
{\cal A}_{\underline{i}} dx^{\underline{i}}
\wedge dt \wedge dx\, ,
\\
& & \\
\hat{A}^{(1)}
&
=
&
-e^{-\frac{4}{3}\hat{\phi}}
\left\{\left(1-e^{\frac{4}{3}\hat{\phi}}\right) dt
+{\textstyle\frac{1}{\sqrt{2}}
{\cal A}_{\underline{i}} dx^{\underline{i}}} \right\}
\\
& & \\
\hat{\phi}
&
=
&
{\textstyle\frac{3}{4}}
\log{(1-{\cal A}_{u})}\, ,
\end{array}
\right.
\end{equation}

\noindent Next, we perform the type--II $T$--duality transformation
Eqs.~(\ref{eq:TAB}) and get the new $GFS(B^\prime)$ solution

\begin{equation}
GFS(B^{\prime})
\left\{
\begin{array}{rcl}
ds^{2}
&
=
&
2e^{-\hat{\varphi}}
\left(du+{\cal A}_{\underline{i}}dx^{\underline{i}} \right)dv
-e^{\hat{\varphi}}dx^{\underline{i}}dx^{\underline{i}}\, ,
\\
& & \\
\hat{\cal B}^{(2)}
&
=
&
e^{-2\hat{\varphi}}
(1-e^{\hat{\varphi}})
{\cal A}_{\underline{i}} dx^{\underline{i}}
\wedge
dv\, ,
\\
& & \\
\hat{\varphi}
&
=
&
\frac{1}{2}\log{(1-{\cal A}_{u})}\, ,
\end{array}
\right.
\end{equation}

\noindent with all other fields vanishing.

Finally, an $xy$--duality transformation (Eqs.~(\ref{eq:xyIIB})) yields
the following $GFS(B)$ solution

\begin{equation}
GFS(B)
\left\{
\begin{array}{rcl}
ds^{2}
&
=
&
2e^{2\hat{\varphi}}
\left(dv+{\cal A}_{\underline{i}}dx^{\underline{i}}\right)du
-dx^{\underline{i}}dx^{\underline{i}}\, ,
\\
& & \\
{\cal{B}}^{(1)}
&
=
&
e^{2\hat{\varphi}}
\left[\left(1-e^{-\frac{1}{2}\hat{\varphi}} \right)dv
+{\cal A}_{\underline{i}}dx^{\underline{i}}\right] \wedge du\, ,
\\
& & \\
\hat{\varphi}
&
=
&
-\frac{1}{2} \log{(1-{\cal A}_{u})}\, .
\\
\end{array}
\right.
\end{equation}

This solution is just the original $GFS$ solution but embedded into the
type--IIB theory.  Therefore, a further type--II $T$--duality
transformation will take us back to the original $SSW$ embedded into the
type--IIA theory, i.e.~the $SSW(A)$ solution we started from.


\subsection{Duality rotation of the GFS}

We next consider the different duality rotations of the $GFS$ solution.
We start from the embedding into the type--IIA theory, i.e.~the $GFS(A)$
solution.  It is given by:

\begin{equation}
GFS(A)
\left\{
\begin{array}{rcl}
ds^{2}
&
=
&
2e^{2\hat{\phi}}
\left(dv + {\cal A}_{\underline{i}}dx^{\underline{i}}\right)du
-dx^{\underline{i}}dx^{\underline{i}}\, ,
\\
& & \\
\hat{B}^{(1)}
&
=
&
2e^{2\hat{\phi}}
\left[\left(1-e^{-2\hat{\phi}}\right)dv +{\cal A}_{\underline{i}}
dx^{\underline{i}} \right]\wedge du\, ,
\\
& & \\
\hat{\phi}
&
=
&
-{\textstyle\frac{1}{2}}\log{(1-{\cal A}_{u})}\, .
\\
\end{array}
\right.
\end{equation}

Performing the $xy$--duality transformations Eqs.~(\ref{eq:xyIIA})
we get the new solution

\begin{equation}
GFS(A^{\prime})
\left\{
\begin{array}{rcl}
ds^{2}
&
=
&
e^{2\hat{\phi}}
\left\{\left(dt +{\textstyle\frac{1}{\sqrt{2}}} {\cal A}_{\underline{i}}
dx^{\underline{i}} \right)^{2} -dx^{2}\right\}
-dx^{\underline{i}} dx^{\underline{i}}\, ,
\\
& & \\
\hat{B}^{(1)}
&
=
&
-e^{2\hat{\phi}}\left[\left( 1- e^{-2\hat{\phi}}\right) dt \wedge dx
+{\textstyle\frac{1}{\sqrt{2}}} {\cal A}_{\underline{i}}
dx^{\underline{i}} \wedge dx \right]\, ,
\\
& & \\
\hat{C}
&
=
&
{\textstyle\frac{\sqrt{2}}{3}}
e^{-2\hat{\phi}}
{\cal A}_{\underline{i}}dx^{\underline{i}} \wedge dt \wedge dx\, ,
\\
& & \\
\hat{\phi}
&
=
&
-{\textstyle\frac{1}{2}} \log{(1-{\cal A}_{u})}\, .
\\
\end{array}
\right.
\end{equation}

\noindent We next apply the type--II $T$--duality rotation
Eqs.~(\ref{eq:TAB}) and get the following $SSW(B^{\prime})$ solution

\begin{equation}
SSW(B^{\prime})
\left\{
\begin{array}{rcl}
ds^{2}
&
=
&
2\left(du +{\cal A}_{u}dv +2{\cal A}_{\underline{i}}dx^{\underline{i}}
\right)dv -dx^{\underline{i}}dx^{\underline{i}}\, ,
\\
& & \\
\hat{B}^{(2)}
&
=
&
2 {\cal A}_{\underline{i}}dx^{\underline{i}}\wedge dv\, ,
\\
& & \\
\hat{\phi}
&
=
&
0\, .
\end{array}
\right.
\end{equation}

\noindent Finally, a further $xy$--duality transformation
Eqs.~(\ref{eq:xyIIB}) gives the solution

\begin{equation}
SSW(B)
\left\{
\begin{array}{rcl}
ds^{2}
&
=
&
2\left(dv +{\cal A}_{u}du +2{\cal A}_{\underline{i}}dx^{\underline{i}}
\right)du -dx^{\underline{i}}dx^{\underline{i}}\, ,
\\
& & \\
\hat{B}^{(1)}
&
=
&
2 {\cal A}_{\underline{i}}dx^{\underline{i}}\wedge du\, ,
\\
& & \\
\hat{\phi}
&
=
&
0\,
\end{array}
\right.
\end{equation}

\noindent which is exactly what one should have expected:
the original $SSW$ solutions embedded into the type--IIB theory.

Note that the above examples do not exhaust the possible new solutions
that can be built out of the $GFS$ and the $SSW$.  It would be of
interest to apply the type--II $S$-- and $T$--dualities to the various
$p$--brane solutions of ten-di\-men\-sion\-al supergravity and to
investigate which solutions are related to each other by some
combination of dualities and which solutions are independent ones.


\subsection{Eleven--dimensional solutions}


We finally consider the lifting of the $SSW$ and $GFS$ solutions to
solutions of the ele\-ven-di\-men\-sion\-al theory.  These liftings lead
to solutions of ele\-ven-di\-men\-sion\-al supergravity which correspond
to a supersymmetric string wave solution and a generalized fundamental
membrane\footnote{The re--interpretation of the ten-di\-men\-sion\-al
string solution as an ele\-ven-di\-men\-sion\-al (extreme) membrane
solution was discussed in \cite{kn:Du3}.  The duality transformations
given in this paper only concern the source--free field equations.  We
will not discuss here the possible source terms and their duality
transformations.} solution which we denote by $SSW_{11}$ and $GFM_{11}$,
respectively.  The explicit form of the $SSW_{11}$ and $GFM_{11}$
solutions is given by:

\begin{equation}
SSW_{11}
\left\{
\begin{array}{rcl}
ds^{2}
&
=
&
2\left(dv +{\cal A}_{u}du +2{\cal A}_{\underline{i}}dx^{\underline{i}}
\right)du -dx^{\underline{i}}dx^{\underline{i}} -dydy\, ,
\\
& & \\
\hat{\hat{C}}
&
=
&
\frac{4}{3} {\cal A}_{\underline{i}}dx^{\underline{i}}\wedge du
\wedge dy\, .
\\
\end{array}
\right.
\label{eq:SSW11}
\end{equation}
and

\begin{equation}
GFM_{11}
\left\{
\begin{array}{rcl}
ds^{2}
&
=
&
(1-{\cal A}_{u})^{-\frac{2}{3}} \left[
2\left(dv   +{\cal
A}_{\underline{i}}dx^{\underline{i}}\right)du -dydy \right]
\\
& & \\
& &
-(1-{\cal A}_{u})^{\frac{1}{3}}dx^{\underline{i}}dx^{\underline{i}}\, ,
\\
& & \\
\hat{\hat{C}}
&
=
&
\frac{4}{3} (1-{\cal A}_{u})^{-1} \left( {\cal A}_u dv + {\cal
A}_{\underline{i}}dx^{\underline{i}}\right)\wedge du \wedge dy\, .
\\
\end{array}
\right.
\label{eq:GFM11}
\end{equation}

We note that the $SSW_{11}$ solution is a generalization of the pp-wave
solution of \cite{kn:Hu2} containing the additional eight functions
${\cal A}_{\underline i}$ while the $GFM_{11}$ solution generalizes the
fundamental membrane solution of \cite{kn:Du2}.  One may verify that the
$SSW_{11}$ given in Eqs.~(\ref{eq:SSW11}) is related to the $GFM_{11}$
solution given in Eqs.~(\ref{eq:GFM11}) by the
ele\-ven-di\-men\-sion\-al type--I $T$--duality rules
Eqs.~(\ref{eq:T11}).  Finally, it would be of interest to apply the
$d=11$ type--I $T$--duality to other ele\-ven-di\-men\-sion\-al
solutions such as the $p$--brane solutions of \cite{kn:Gu1}.

\section*{Acknowledgements}

One of us (E.B.) would like to thank the physics department of QMW
college for its hospitality.  The work of E.B.~has been made possible by
a fellowship of the Royal Netherlands Academy of Arts and Sciences
(KNAW).  This work was supported by European Union {\sl Human Capital
and Mobility} program grants.

\appendix


\section{Conventions}
\label{sec-conventions}

We use double hats for ele\-ven-di\-men\-sion\-al objects, single hats
for ten-di\-men\-sion\-al objects and no hats for
ni\-ne-di\-men\-sion\-al objects.  Greek or underlined indices are world
indices, and latin or non-underlined indices are Lorentz indices.  We
use the indices $\hat{\hat{\mu}}=(\hat{\mu},y)=(\mu,x,y)$, with
$y=x^{\underline{10}}$ and $x=x^{\underline{9}}$.  Our signature is
$(+--\ldots-)$.  The antisymmetric Levi-Civita tensor ${\hat{\hat
\epsilon}}$ is defined by

\begin{equation}
\hat{\hat{\epsilon}}^{\hat{\hat{\mu}}_{0}\ldots
\hat{\hat{\mu}}_{10}} = 1.
\end{equation}

Our spin connection $\omega$ (in $d$ dimensions) is defined by

\begin{equation}
\omega_{\mu}{}^{ab}(e)= -e^{\nu[a}
\left( \partial_{\mu}e_{\nu}{}^{b]} -\partial_{\nu}e_\mu{}^{b]}
\right) - e^{\rho[a} e^{\sigma b]}
\left( \partial_{\sigma} e_{c\rho} \right) e_{\mu}{}^{c}\ .
\end{equation}

\noindent The curvature tensor corresponding to this spin connection
field is defined by

\begin{equation}
R_{\mu\nu}{}^{ab}(\omega)=
2\partial_{[\mu}\omega_{\nu]}{}^{ab}-
2\omega_{[\mu}{}^{ac}
\omega_{\nu]c}{}^b\, ,
\hskip 1truecm
R(\omega)\equiv
e^{\mu}{}_{a}e^{\nu}{}_{b} R_{\mu\nu}{}^{ab}(\omega)\ .
\end{equation}

Although we don't use differential forms, sometimes we use the following
convention: when indices are not shown explicitly, we assume that all of
them are world indices and all of them are completely antisymmetrized in
the obvious order.  For instance

\begin{equation}
\hat{G} = \partial\hat{C} -2\hat{H}^{(1)}\hat{A}^{(1)}\, ,
\end{equation}

\noindent means

\begin{equation}
\hat{G}_{\hat{\mu}\hat{\nu}\hat{\sigma}\hat{\rho}} =
\partial_{[\hat{\mu}} \hat{C}_{\hat{\nu}\hat{\rho}\hat{\sigma}]}
-2\hat{H}^{(1)}_{[\hat{\mu}\hat{\nu}\hat{\rho}}
\hat{A}^{(1)}_{\hat{\sigma}]}\, .
\end{equation}


\section{Eleven-- And Nine--dimensional Fields}
\label{sec-11vs9}

Here we present the expression of the ele\-ven-di\-men\-sion\-al fields
in terms of the ni\-ne-di\-men\-sion\-al ones.  The components of the
ele\-ven-di\-men\-sion\-al metric are

\begin{equation}
\begin{array}{rclrcl}
\hat{\hat{g}}_{\underline{y}\underline{y}} & = &
-k^{\frac{2}{3}} e^{\frac{4}{3}\phi}\, , &
\hat{\hat{g}}_{\underline{x}\underline{x}} & = &
-k^{\frac{2}{3}} e^{\frac{4}{3}\phi}
\left(\ell^{2} +ke^{-2\phi}\right)\, ,
\\
& & & & &
\\
\hat{\hat{g}}_{\underline{y}\underline{x}} & = &
-\ell k^{\frac{2}{3}} e^{\frac{4}{3}\phi}\, , &
\hat{\hat{g}}_{\mu\underline{y}} & = &
-k^{\frac{2}{3}} e^{\frac{4}{3}\phi} A_{\mu}^{(1)}
-\ell k^{\frac{2}{3}} e^{\frac{4}{3}\phi} A_{\mu}^{(2)}\, ,
\\
& & & & &
\\
\hat{\hat{g}}_{\mu\underline{x}} & = &
-\ell k^{\frac{2}{3}} e^{\frac{4}{3}\phi} A_{\mu}^{(1)}
-k^{\frac{2}{3}} e^{\frac{4}{3}\phi}
\left(\ell^{2} +ke^{-2\phi} \right) A_{\mu}^{(2)}\, ,
\hspace{-3cm}
& & &
\\
& & & & &
\\
\hat{\hat{g}}_{\mu\nu} & = &
-k^{-\frac{1}{3}} e^{-\frac{2}{3}\phi} g_{\mu\nu}
-k^{\frac{2}{3}} e^{\frac{4}{3}\phi} A_{\mu}^{(1)} A_{\nu}^{(1)}
-k^{\frac{2}{3}} e^{\frac{4}{3}\phi}
\left(\ell^{2} +ke^{-2\phi}\right) A_{\mu}^{(2)} A_{\nu}^{(2)}
\hspace{-10cm} & & &
\\
& & & & &
\\
& &
-2\ell k^{\frac{2}{3}} e^{\frac{4}{3}\phi}
A_{\mu}^{(1)} A_{\nu)}^{(2)}\, ,
& & &
\\
\end{array}
\end{equation}

\noindent and the components of the ele\-ven-di\-men\-sion\-al
three-form $\hat{\hat{C}}$ are

\begin{equation}
\begin{array}{rclrcl}
\hat{\hat{C}}_{\mu\nu\rho} & = & C_{\mu\nu\rho}\, , &
\hat{\hat{C}}_{\mu\underline{x}\underline{y}} & = &
-\frac{2}{3} B_{\mu}\, ,
\\
& & & & &
\\
\hat{\hat{C}}_{\mu\nu\underline{y}} & = &
\frac{2}{3}\left(B^{(1)}_{\mu\nu}
+A^{(2)}_{[\mu}B_{\nu]}\right)\, ,&
\hat{\hat{C}}_{\mu\nu\underline{x}} & = &
\frac{2}{3}\left(B^{(2)}_{\mu\nu}
-A^{(1)}_{[\mu}B_{\nu]}\right)\, . \\
\end{array}
\end{equation}

The inverse relations are

\begin{equation}
\begin{array}{rclrcl}
k & = & \left( -\hat{\hat{g}}_{\underline{y}\underline{y}}
\right)^{-\frac{1}{4}} \Delta^{\frac{1}{2}}\, , &
A_{\mu}^{(2)} & = & \left(\hat{\hat{g}}_{\underline{x}\mu}
\hat{\hat{g}}_{\underline{y}\underline{y}}
-\hat{\hat{g}}_{\underline{x}\underline{y}}
\hat{\hat{g}}_{\mu\underline{y}}
\right)/\Delta\, ,
\\
& & & & &
\\
\ell & = & \hat{\hat{g}}_{\underline{x}\underline{y}}
/\hat{\hat{g}}_{\underline{y}\underline{y}}\, , &
A_{\mu}^{(1)} & = & \left(\hat{\hat{g}}_{\underline{y}\mu}
\hat{\hat{g}}_{\underline{x}\underline{x}}
-\hat{\hat{g}}_{\underline{x}\underline{y}}
\hat{\hat{g}}_{\mu\underline{x}}
\right)/\Delta\, ,
\\
& & & & &
\\
\phi & = & \frac{1}{8}\log{\left[\left(
-\hat{\hat{g}}_{\underline{y}\underline{y}} \right)^{7}
/\Delta^{2} \right]}\, , &
B_{\mu} & = & \frac{3}{2}
\hat{\hat{C}}_{\underline{x}\mu\underline{y}}\, ,
\\
& & & & &
\\
B^{(1)}_{\mu\nu} & = & \frac{3}{2}\left[
\hat{\hat{C}}_{\mu\nu\underline{y}}\Delta
+(\mu \leftrightarrow \underline{x}) \right]/\Delta\, , &
B^{(2)}_{\mu\nu} & = & \frac{3}{2}\left[
\hat{\hat{C}}_{\mu\nu\underline{x}}\Delta
+(\mu \leftrightarrow \underline{y}) \right]/\Delta\, .
\\
\end{array}
\end{equation}

\noindent where

\begin{equation}
\Delta = \hat{\hat{g}}_{\underline{x}\underline{x}}
\hat{\hat{g}}_{\underline{y}\underline{y}}
-\hat{\hat{g}}_{\underline{x}\underline{y}}^{2}\, .
\end{equation}

The expression of $g_{\mu\nu}$ in terms of the ele\-ven-di\-men\-sion\-al
fields is not very enlightening and, in any, case, it can be readily
obtained from the above formulae.


\end{document}